\documentclass[12pt]{iopart}

\usepackage{graphicx}
\usepackage{subcaption}
\usepackage{siunitx}
\usepackage{longtable,tabularx}
\usepackage{xcolor}
\usepackage[top=25truemm,bottom=20truemm,left=20truemm,right=20truemm]{geometry}
\usepackage{setspace}
\usepackage{multirow}
\usepackage{scalerel}
\usepackage{cite}
\usepackage[normalem]{ulem}

\begin{document}

\title[Keita Nishii PSST 2023]{Kinetic Simulation of Ion Thruster Plume Neutralization in a Vacuum Chamber}

\author{Keita Nishii and Deborah A. Levin}

\address{104 S Wright St, Urbana, Illinois, US}
\ead{knishii@illinois.edu}
\vspace{10pt}
\begin{indented}
\item[] \today
\end{indented}

%
%
%
%
%




\begin{abstract}
The electrical environment of a ground vacuum testing chamber creates facility effects for gridded ion thrusters. For example, it is well known that the plume from the thruster generates current paths that are very different from what occurs in space, and the neutralization of this plume is also different. For reasons such as this, it is important to clarify how the experimental testing environment affects plasma flows, but understanding this effect solely through ground experiments is difficult. To that end, this study utilizes particle-in-cell and direct simulation Monte Carlo methods to simulate xenon beam ions and electrons emitted from a neutralizer. First, we compare simulations conducted within the chamber to those conducted in space, demonstrating that grounded chamber walls increase the electric potential and electron temperature. Next, we investigate the impact of the neutralizer's position and the background pressure on the plume in the vacuum chamber.  We find that as the neutralizer position moves closer to the location of maximum potential, more electrons are extracted, resulting in increased neutralization of the plume. We also observe that high background pressure generates slow charge-exchange ions, creating ion sheaths on the side walls that alter ion current paths. Finally, we discuss how the potential at the thruster and neutralizer exits affects the plume. The relative potential of the neutralizer to the vacuum chamber wall is observed to significantly influence the behavior of the electrons, thereby altering the degree of plume neutralization. These findings are shown to be consistent with experimental results in the literature and demonstrate the promise of high-performance simulation.
\end{abstract}

\section{Introduction}

Gridded ion thrusters (GITs), an electric propulsion device commonly used due to its high-specific impulse ($>$ 3000 s)~\cite{Holste2020-gr}, generate thrust by selectively extracting ions using ion optics (grids). However, the performance of GITs relies not only on the ion optics but also on the neutralizer attached near the ion source location. Thermal electrons are released from the neutralizer to mitigate spacecraft charging caused by the space-charge effect of excess ion emission from the thruster, and the coupling between the two sources greatly affects the performance of the neutralizer. Research on this coupling of neutralizers dates back to the SERT II satellite~\cite{Kerslake1993-my}, which demonstrated that the operation of the neutralizer reduced the electric potential of the ion plume and suppressed the decrease in spacecraft floating potential in space~\cite{Kerslake1982-mj}. Nakayama et al.~\cite{Nakayama2015-ew} also investigated the ion current in a ground vacuum chamber by changing the neutralizer electron current and revealed that more ions returned to the thruster exit and body when the neutralizer electron current was insufficient.

Past GIT plume experiments have mainly focused on beam ion current and energy, which directly relate to thrust, unlike Hall effect thrusters, where the electron transport from the external neutralizer is important for plasma generation~\cite{Dale2020-dd, Li2019-eo}. In addition to current and energy, Polansky et al.~\cite{Polansky2013-nv} studied ion and electron number densities downstream in the radial direction, and Conde et al.~\cite{Conde2022-zp} investigated the energy distribution of ions and electrons in a GIT plume. However, to date, few studies have experimentally investigated the detailed electron motion.

GIT plume studies are extensively based on numerical calculations, with most assuming quasi-neutrality or being solved by fluid models to simplify the electrodynamics treatment. However, full kinetic simulations, in which both electrons and ions are modeled as macroparticles, are essential to understanding electron transport. Table~\ref{tab:paststudies} summarizes previous studies that simulated GIT plumes using a fully kinetic approach~\cite{Zhao2018-kq, Hu2015-mr, Wang2012-hy, Hu2020-vs, Wang2015-pm, Li2019-xn, Wang2019-gy, Nuwal2020-pa, Jambunathan2020-io, Nishii2023-ju, Usui2013-al, Jambunathan2018-xf, Brieda2018-wj, Brieda2005-kj, Jambunathan2020-yx}. Among them, Refs~\cite{Nuwal2020-pa, Jambunathan2020-io, Nishii2023-ju, Jambunathan2018-xf, Jambunathan2020-yx} combine an explicit particle-in-cell (PIC)~\cite{Birdsall1991-lh} with direct simulation Monte Carlo (DSMC)~\cite{Bird1987-ba} to solve charge-exchange (CEX) and momentum-exchange (MEX) collisions between ions and neutral particles. In addition, Refs~\cite{Hu2015-mr, Brieda2018-wj, Wang2019-gy, Jambunathan2020-io, Nuwal2020-pa} have shown that the fluid approximation of the electrons used with the Boltzmann or polytropic relations cannot adequately simulate the plume potential and electron density. 

\begin{table}[hbt!]
\caption{Previous study of the gridded ion thruster plume using a fully kinetic PIC approach.}
\centering
\small
\begin{tabular}{llccccl}
\hline \hline
Author [Ref.]             & Neutralizer pos. & Species & State     & $R_0/\lambda_\mathrm{D0}^a$ & Neutral col. & Geometry \\  \hline
Zhao~\cite{Zhao2018-kq}   & Co-located           & Proton  & Transient & 5             & No      & In-space \\ 
Hu~\cite{Hu2015-mr}       & Co-located           & Proton  & Transient & 20            & No      & In-space \\ 
Wang~\cite{Wang2012-hy}   & Co-located           & Proton  & Steady    & 20            & No      & In-space \\
Hu~\cite{Hu2020-vs}       & Co-located           & Proton  & Steady    & 50            & CEX$^b$ & In-space \\
Wang~\cite{Wang2015-pm}   & Co-located           & Proton  & Steady    & 7             & CEX$^b$ & In-chamber \\
Li~\cite{Li2019-xn}       & Co-located           & Proton  & Steady    & 10            & No      & In-space \\
Wang~\cite{Wang2019-gy}   & Co-located           & Xenon   & Steady    & 20            & No      & In-space \\
Nuwal~\cite{Nuwal2020-pa} & Co-located           & Xenon   & Steady    & 120           & CEX/MEX$^c$ & In-space w/SAP$^d$ \\
Jabunathan~\cite{Jambunathan2020-io} & Co-located & Xenon  & Steady    & 285           & CEX/MEX$^c$ & In-space w/SAP$^d$ \\
Nishii~\cite{Nishii2023-ju} & Co-located         & Xenon   & Steady    & 120           & CEX/MEX$^c$ & In-chamber \\ \hline
Usui~\cite{Usui2013-al}   & External             & Proton  & Transient & 20            & No      & In-space \\ 
Jabunathan~\cite{Jambunathan2018-xf} & External & Xenon & Transient & 19    & No      & In-space \\ 
Brieda~\cite{Brieda2018-wj} & Internal & Oxygen & Steady    & 3             & No      & In-space \\ 
Brieda~\cite{Brieda2005-kj} & External            & Xenon  & Steady    & 10        & No      & In-space \\
Jabunathan~\cite{Jambunathan2020-yx} & External & Xenon & Steady    & 19    & No      & In-space \\ 
{\bf This study}          & External              & Xenon  & Steady    & 19            & CEX/MEX$^c$ & {\bf In-chamber} \\ 
\hline \hline
\multicolumn{6}{l}{$^a$ $R_0$ is radius of an ion source, and $\lambda_{D0}$ is the reference Debye length.} \\
\multicolumn{6}{l}{$^b$ Calculated by Monte Carlo Collisions calculation. $^c$ Calculated by DSMC calculation.} \\
\multicolumn{6}{l}{$^d$ SAP means a solar array panel geometry.} \\
\end{tabular}
\label{tab:paststudies}
\end{table}

The previous studies shown in Table~\ref{tab:paststudies} can be divided into two main categories regarding the neutralizer position. References~\cite{Zhao2018-kq, Hu2015-mr, Wang2012-hy, Hu2020-vs, Wang2015-pm, Li2019-xn, Wang2019-gy, Nuwal2020-pa, Jambunathan2020-io, Nishii2023-ju} placed the ion and electron sources at the same location (co-located position), with the plume immediately neutralized just downstream from the thruster exit. In actual GITs, however, a large potential gradient occurs in principle because the ions and electrons are supplied from separate positions. Other work listed in Table~\ref{tab:paststudies}, Refs.~\cite{Usui2013-al, Jambunathan2018-xf, Brieda2005-kj, Brieda2018-wj, Jambunathan2020-yx}, studied plume neutralization for such shifted electron source position cases as described "external" or "internal" in the table. It was observed that the plume potential significantly changed when comparing co-located and external source cases, even for the same electron source size and density~\cite{Jambunathan2018-xf, Jambunathan2020-yx}. The external source cases have relatively smaller $R_0/\lambda_\mathrm{D0}$ than the co-located cases, 
where $R_0$ and $\lambda_\mathrm{D0}$ are the thruster exit radius and the initial Debye length, respectively. This means that they target smaller thrusters or lower-density plasmas because the electron density at the shifted neutralizer is normally much larger than the ion beam density. Hence, the minimum mesh size becomes smaller than the co-located cases, increasing computational costs.

Table~\ref{tab:paststudies} also shows that many simulations model space operations~\cite{Zhao2018-kq, Hu2015-mr, Hu2020-vs, Wang2015-pm, Li2019-xn, Wang2019-gy, Nuwal2020-pa, Jambunathan2020-io, Usui2013-al, Jambunathan2018-xf, Brieda2005-kj, Brieda2018-wj, Jambunathan2020-yx}. Although the in-space environment has almost an infinite volume for plume expansion, an otherwise-trapped electron may reach the end of the computational domain before it can reverse direction due to finite computational resources. Thus, Refs.~\cite{Li2019-xn, Brieda2018-wj, Jambunathan2020-yx} have focused on developing electron boundary conditions for fully kinetic simulations. As can be seen from the table, however, few studies have been conducted on ground chamber tests. 

Ground tests of electric propulsion systems can introduce significant uncertainty in the on-orbit performance prediction because of facility effects~\cite{Foster2022-ck}. For example, Nishii et al.~\cite{Nishii2023-ju} simulated the contamination caused by carbon backsputtering for different sputter models and plume conditions and demonstrated the importance of combining backsputtering and PIC plume simulations. Hu et al.\cite{Hu2020-vs} studied electrical effects by simulating a proton ion beam with different beam radii and showed that the beam width became 20 times larger than the thruster radius where ions reached their maximum velocity. Their work suggested that the vacuum chamber facility may significantly affect the plasma beam by prematurely terminating its expansion if the chamber size is less than 20 times the beam radius. However, since xenon has a higher mass than that of a proton (their assumed ion mass), the plume diffusion in their simulation would be smaller than that of an actual GIT. In addition, since the ion and electron sources were in the same position, which is not realistically possible, the effect of coupling of distant sources was neglected. Therefore, the electrical effects that affect the prediction of the neutralizer coupling voltage, the neutralizer bias voltage from the thruster common, and the plume divergence have not been fully investigated. 

The primary objective of this study is to investigate electrical facility effects related to GITs using a fully kinetic simulation. We use an in-house developed 3-D PIC-DSMC solver, Cuda-based Hybrid Approach for Octree Simulations (CHAOS)~\cite{Jambunathan2018-xf}, which enables us to solve for a steady-state GIT plume with relatively large $R_0/\lambda_\mathrm{D0}$ for the true mass ratio between xenon ions and electrons, as shown in Table~\ref{tab:paststudies} (Refs.~\cite{Nuwal2020-pa, Jambunathan2020-io, Nishii2023-ju, Jambunathan2018-xf, Jambunathan2020-yx}). This study specifically seeks to understand the effect of the vacuum chamber boundary condition, the neutralizer exit location, and the finite background neutral density on the GIT plume, including beam ions and neutralizer electrons.

Additionally, the biased (non-zero) potential that appears in the ground chamber during GIT operation has not previously been investigated in numerical simulations. For instance, in a two-grid GIT system, there is no deceleration grid (decel grid), which is used to reduce the CEX ion backflow that can cause acceleration grid (accel grid) erosion~\cite{Wirz2008-fj}. In such cases, the accel grid at a negative voltage is exposed directly to the plasma plume. Second, the neutralizer coupling voltage is normally altered to increase the neutralizer current~\cite{Kerslake1993-my, Nakayama2015-ew, Polansky2013-nv}. Taking into account the keeper positive voltage~\cite{Kerslake1993-my}, which is crucial to the extraction of electrons from the hollow cathode, the neutralizer exit voltage should differ from the thruster common voltage. Therefore, we also investigate how such electrical potential boundaries affect the GIT plume in the vacuum chamber.

The outline of the remainder of this article is as follows. Section~\ref{sec:numericalapproach} reviews our plasma modeling approach and describes the boundary conditions implemented in CHAOS. Section~\ref{sec:conditions} explains the geometry, species, and numerical conditions of the study and selection of variable parameters. Finally, we present and discuss comparisons between the simulated cases, including the effect of the simulation boundary condition (Section~\ref{sec:result-spchamber}), neutralizer locations (Section~\ref{sec:result-neutpos}), background neutral particles (Section~\ref{sec:result-CEX}), and electric potential at the thruster (Section~\ref{sec:result-decelgrid}) and neutralizer (Section~\ref{sec:result-keepervoltage}) exits.


\section{Numerical Approach}
\label{sec:numericalapproach}

\subsection{PIC and DSMC Modules and Their Coupling}
\label{subsec:dsmc-pic}

In this section, we briefly discuss the computational framework implemented in CHAOS to couple the PIC and DSMC approaches in order to calculate the self-consistent electric field, taking into account the reaction between ions and neutral particles. In the EP plume, collisions and electric fields operate on significantly different time and length scales, differing by at least two orders of magnitude. To mitigate the impact of these differences, CHAOS employs several computational techniques as previously described in our earlier papers~\cite{Nuwal2020-pa, Jambunathan2020-io, Nishii2023-ju, Jambunathan2018-xf, Jambunathan2020-yx}.

The DSMC module models three types of collisions: momentum exchange (MEX) collisions between Xe-Xe and Xe-Xe$^+$, and CEX collisions between Xe-Xe$^+$. The collision cross sections for MEX between neutral particles and MEX and CEX collisions between neutral particles and ions are obtained from Refs.~\cite{Araki2013-em}, and \cite{Miller2002-ne}. The no-time-counter collision scheme proposed by Serikov et al.~\cite{Serikov1999-jq} is used in this study since it accounts for the disparate timesteps and weighting factors of ions and neutral particles. The neutral particles move only when the DSMC module is executed, every 100 PIC timesteps, while the ions and electrons move every iteration.

In the PIC module, the electric potential is calculated using an explicit PIC technique. In the fully kinetic approach, the electric field, $E$, is self-consistently solved by:
\begin{equation}
    \label{eq:chargedens}
    \rho=e\left(n_\mathrm{i}-n_\mathrm{e}\right),
\end{equation}
\begin{equation}
    \label{eq:poisson}
    \nabla^2\phi = - \frac{\rho}{\epsilon_0},
\end{equation}
\begin{equation}
    \label{eq:efield}
    E = - \nabla\phi,
\end{equation}
where $\rho$ is the charge density, $e$ is the elementary charge, $n_\mathrm{i}$ and $n_\mathrm{e}$ are the number density of ions and electrons, $\phi$ is the electric potential, and $\epsilon_0$ is the permittivity of free space. A finite volume approach based on an unstructured octree grid is used to solve Eq.~(\ref{eq:poisson}).

CHAOS has a number of PIC and DSMC coupling algorithms that save computational effort. First, using a Morton Z-curve, CHAOS constructs two separate grids with a linearized forest of octrees (FOT) in the PIC and DSMC modules, respectively, because the mean free path, $\lambda$, and Debye length, $\lambda_D$, differ by at least three orders of magnitude. The FOT for DSMC (C-FOT) is constructed to resolve the local mean free path, while the FOT for PIC (E-FOT) is constructed to resolve the local Debye length, where the refined cell size $x <\lambda_D$. We apply an adaptive mesh refinement method since the number density can vary widely in the computational domain. Both C- E-FOTs are reconstructed every 20,000 iterations before sampling starts. Second, weighting factors, $W$, are utilized to increase the number of charged computational particles compared to the neutral particles due to the disparate length scales of the C- and E-FOTs and disparate number densities of the neutral particles and CEX ions. In this study, the ratio of neutral and ion ($W_\mathrm{n}/W_\mathrm{i}$) is set at 200,000. Third, time-slicing of the DSMC and PIC modules and species-dependent timesteps are implemented due to the different timescales for collision and plasma frequencies. The positions of the neutral particles, ions, and electrons are updated with timesteps of $\Delta t_\mathrm{n} \gg \Delta t_\mathrm{i} = \Delta t_\mathrm{e}$ to reconcile these disparate timescales. In this study, we use a timestep of $\Delta t_\mathrm{n} = 1.12 \times 10^{-4}$ s for neutral particles and $\Delta t_\mathrm{i,e} = 2.8 \times 10^{-10}$ s for both ions and electrons.

\subsection{Boundary Conditions}
\label{subsec:bcsetting}

To satisfy the objectives of this study, we use both in-space and in-chamber boundary conditions (BCs) for the outer edge domain boundary. For the {\it in-space simulation}, the charge-conserving energy-based BC (CCE BC) developed by Jambunathan and Levin~\cite{Jambunathan2020-yx} is used for the downstream boundary ($z=0.8$ m), and the buffer BC is used for the other boundaries to simulate the infinite expansion of the thruster plume, similar to our previous calculations (Refs~\cite{Nuwal2020-pa, Jambunathan2020-yx}). The buffer BC simulates the inflow of electrons from outside by placing a buffer region outside the computational boundary and copying the particles inside the boundary out to a distance of $\lambda_\mathrm{D0}$ beyond it. The CCE BC specularly reflects some electrons arriving at the edge of the computational domain and eliminates others using the following approach. The baseline total charge, $Q_0$, and the average electron kinetic energy in the computational domain, $\langle E_\mathrm{e} \rangle$, are obtained at the time step just before the beam ions reach the downstream domain boundary. In subsequent timesteps, when the total charge in the domain is less than $Q_0$, electrons with energies less than $\langle E_\mathrm{e} \rangle$ are specularly reflected. In Ref.~\cite{Jambunathan2020-yx}, it was verified that these BCs satisfy the requirements needed for the plume modeling by changing the size of the domain (see Table II in Ref.~\cite{Jambunathan2020-yx}). In terms of electrical boundary conditions, the inhomogeneous Neumann BC for the electric potential is implemented on all domain boundaries in the in-space simulations. For each boundary, the normal potential gradient $({\partial \phi}/{\partial n})_\mathrm{bc}$ is computed based on the current density through the boundary as follows; 
\begin{equation}
    \left(\frac{\partial \phi}{\partial n}\right)_\mathrm{bc} = \frac{e (N_\mathrm{i,bc}-N_\mathrm{e,bc})}{A_\mathrm{bc}\epsilon_0},
\end{equation}
where $N_\mathrm{i,bc}$ and $N_\mathrm{e,bc}$ are the number of ions and electrons that cross the boundary, and $A_\mathrm{bc}$ is the area of the boundary.

For the {\it in-chamber} boundary conditions, we use a fully diffuse reflection condition with a 300~K accommodation for ions and neutral particles on the chamber walls at the edge of the computational domain. The walls absorb all electrons by removing them from the domain and neutralize all incident ions by returning them into the domain as neutral particles with a temperature of 300~K. The 0~V Dirichlet BC is implemented on every boundary surface in the PIC module. The plasma screen, which is the housing of the thruster assembly and is normally electrically grounded, has the same BC as the vacuum chamber walls. We also assign a charge-absorbing BC for particles impinging on the thruster and neutralizer exits 
 and a Dirichlet potential BC for the electric potential with a baseline value of 0 V. The detailed settings about the potential boundaries are described in Section~\ref{subsec:casedescription}.


\section{Simulation Settings}
\label{sec:conditions}

\subsection{Calculation Geometry and Species}
\label{subsec:geomandspecies}

\begin{figure}[hbt!]
    \begin{subfigure}{0.355\textwidth}
        \centering
        \includegraphics[width=\linewidth]{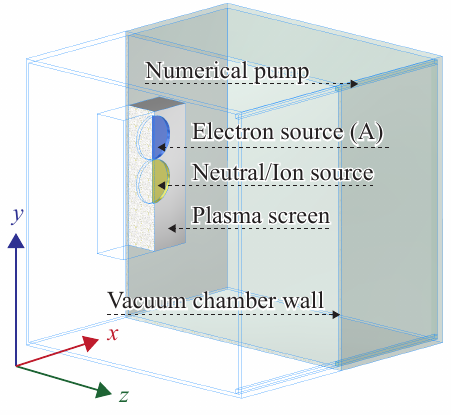}
        \caption{Three dimension} 
        \label{fig:abreast3d}
    \end{subfigure}%
    \begin{subfigure}{0.245\textwidth}
        \centering
        \includegraphics[width=\linewidth]{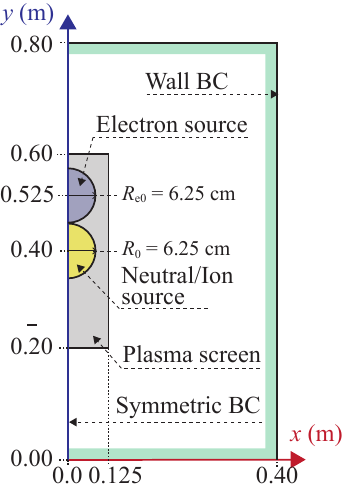}
        \caption{$x$-$y$ plane} 
        \label{fig:abreastxy}
    \end{subfigure}%
    \begin{subfigure}{0.4\textwidth}
        \centering
        \includegraphics[width=\linewidth]{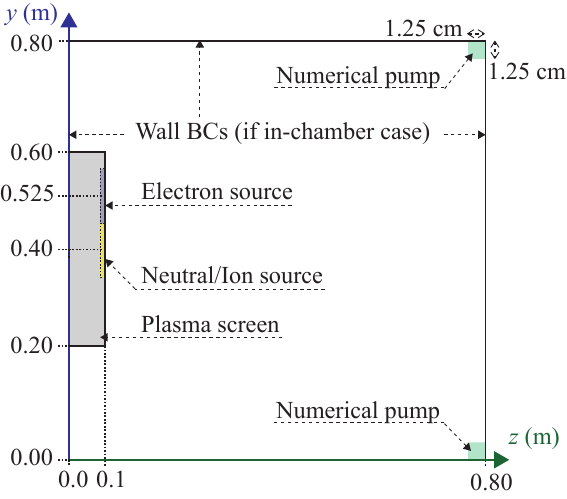}
        \caption{$z$-$y$ plane} 
        \label{fig:abreastzy}
    \end{subfigure}%
    \caption{Computational domain setups for Type-A electron source cases, where the thruster exit radius, $R_\mathrm{0}$, and the neutralizer exit radius, $R_\mathrm{e0}$, are the same as $R_\mathrm{0} = R_\mathrm{e0} = 6.25$ cm.}
    \label{fig:Abreastschem}
\end{figure}

\begin{figure}[hbt!]
    \begin{subfigure}{0.355\textwidth}
        \centering
        \includegraphics[width=\linewidth]{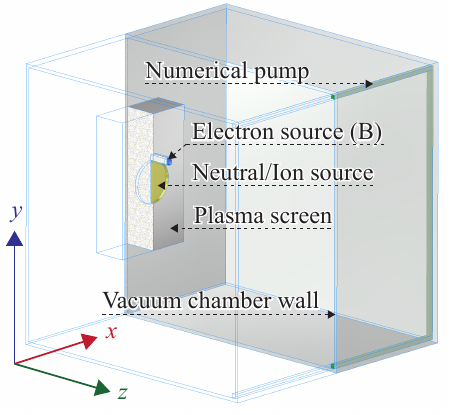}
        \caption{Three dimension} 
        \label{fig:downstream3d}
    \end{subfigure}%
    \begin{subfigure}{0.245\textwidth}
        \centering
        \includegraphics[width=\linewidth]{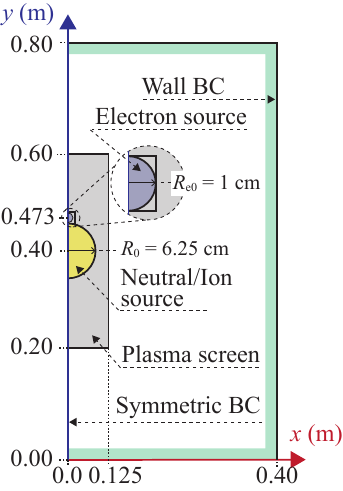}
        \caption{$x$-$y$ plane} 
        \label{fig:downstreamxy}
    \end{subfigure}%
    \begin{subfigure}{0.4\textwidth}
        \centering
        \includegraphics[width=\linewidth]{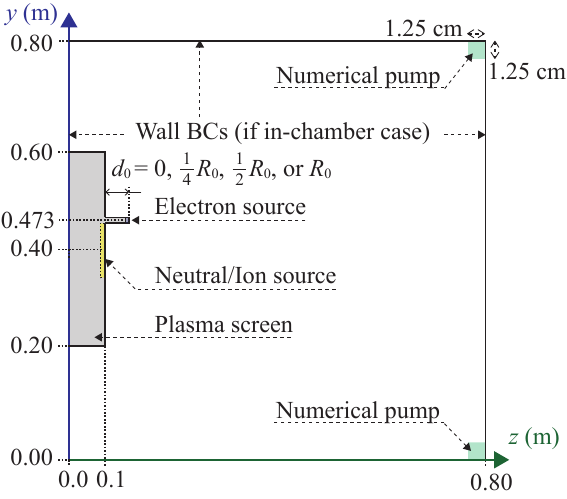}
        \caption{$z$-$y$ plane} 
        \label{fig:downstreamzy}
    \end{subfigure}%
    \caption{Computational domain setups for Type-B electron source cases, where the neutralizer exit radius, $R_\mathrm{e0}=1$ cm, is smaller than the thruster exit radius, $R_\mathrm{0}=6.25$ cm. The distance between the neutralizer exit and thruster body, $d_0$, changes for four cases as $d_0 = 0, \frac{1}{4}R_\mathrm{0}, \frac{1}{2}R_\mathrm{0}$ or, $R_\mathrm{0}$.}
    \label{fig:Downstreamschem}
\end{figure}

Figures~\ref{fig:Abreastschem} and \ref{fig:Downstreamschem} show the three- and two-dimensional schematics of the in-chamber cases investigated in this study. Note that only the geometry of the neutralizer differs in Figs.~\ref{fig:Abreastschem} and \ref{fig:Downstreamschem}. The GIT is placed in a cubic vacuum chamber with a length of 0.8~m per side. In this study, only a half of the domain is simulated due to symmetry to save computational effort, i.e., a specular reflection BC and a Neumann BC ($\partial \phi/ \partial n=0$) are implemented on the $x=0$~m plane. Numerical pumps, shown as green volumes, are installed at all corners of the downstream face to remove heavy neutral particles from the vacuum chamber. Computational particles entering the numerical pump volume are deleted from the calculation, utilizing the same method employed in our previous studies~\cite{Korkut2017-zq, Nishii2023-ju}. The cross-sectional area of the numerical pump is assumed to be $1.25 \times 1.25$ cm$^2$, which produces a typical vacuum chamber pressure of $10^{-6}$ Torr.

The dimension of the thruster, which emits Xe neutrals and Xe$^+$ ions, is the same as that used in our previous calculations of an ion thruster system~\cite{Nuwal2020-pa, Jambunathan2020-io, Nishii2023-ju, Jambunathan2020-yx}. The center position of the cylindrical thruster exit with a radius, $R_0$, of 0.0625~m is located at ($x_0, y_0, z_0$) = (0.0, 0.4, 0.1) m. The domain contains the plasma screen offset from the inlet plane at $z = 0.1$ m. This study examines two neutralizer sizes and number densities. The first configuration is designated as "A," which has the same characteristics as in our previous work~\cite{Jambunathan2020-io}, as illustrated in Fig~\ref{fig:Abreastschem}; (a): three-dimensional view, (b): two-dimensional view seen in the $x$-$y$ plane, and (c): in the $z$-$y$ plane. This neutralizer has the same radius as the thruster and is placed side by side on the same plane. In addition to the type A configuration, this study analyzes a more realistic configuration designated as "B." This neutralizer has a higher electron density and a smaller outlet downstream from the thruster, as shown in Fig~\ref{fig:Downstreamschem}; (a): three-dimensional view, (b): two-dimensional view seen in the $x$-$y$ plane, and (c): in the $z$-$y$ plane. The neutralizer exit radius, $R_\mathrm{e0}$, is set to 1 cm based on the size of a typical neutralizer~\cite{Kitamura2007-sk} and the exit electron number density is 32 times greater than type A to obtain the same current. Unlike the type A cases, the neutralizer exit shifts far from the thruster exit by a value of $d_0$ in the $z$-direction to investigate the effect of the electron exit position on the plume.

Table~\ref{tab:thrustercondition} summarizes the conditions of each species at the thruster and neutralizer exits. Similar to previous mesothermal studies~\cite{Wang2012-hy, Jambunathan2018-xf, Jambunathan2020-yx}, we chose a ratio of the initial ion temperature, $T_\mathrm{i0}$, to the initial electron temperature, $T_\mathrm{e0}$, of 0.01. All species are initialized at their sources with a stationary half-Maxwellian distribution in the streamwise and full-Maxwellian in the cross-stream directions. The reference plasma number density is considered to be the same as the ion number density at the thruster exit of $n_\mathrm{0} = 1.0 \times 10^{13}$ \si{m^{-3}}. These selected values for $T_\mathrm{e0}$ and $n_\mathrm{0}$ result in an initial Debye length, $\lambda_\mathrm{D0} = 3.32 \times 10^{-3}$~m, initial electron plasma frequency, $\omega_\mathrm{pe} = 1.78 \times 10^8$ rad/s, and initial electron thermal velocity, $v_\mathrm{te0} = 592,892$ m/s. In the kinetic simulations, the ion and electron timesteps should follow the requirements of $\Delta t < 0.1 \omega_\mathrm{pe}$. In this work, we use a timestep of $\Delta t = 0.05 \omega^{-1}_\mathrm{peo} = 2.8 \times 10^{-10}$ s for ions and electrons. The superparticle parameter, $F_\mathrm{num}$, is set at 2,500 for all simulations such that there are at least 15 particles per species per cell for both the C-FOT and E-FOT. As a result, the total number of computational ion and electron particles is about 20 million at steady state. Table~\ref{tab:thrustercondition} also gives the Xe neutral particle parameters for the cases with a background pressure; otherwise, the plume is assumed to be collisionless. The exit number density of neutrals is $n_\mathrm{n0}=1.0\times 10^{17}$ m$^{-3}$, which is the typical order for actual GITs giving a total number of computational neutral particles of about five million at steady state. 

\begin{table}[hbt!]
\caption{Parameters of the species at the thruster and neutralizer exits.}
\centering
\small
\begin{tabular}{lcccc}
\hline \hline
Thruster exit conditions  & Xe  & Xe$^+$ & e$^-$ (Type-A) & e$^-$ (Type-B) \\ \hline
Source center$^a$ (m)     & (0.0, 0.4, 0.1) & (0.0, 0.4, 0.1) & (0.0 ,0.525, 0.1) & (0.0, 0.4725, 0.1+$d_0$) \\
Source radius (cm)        & 6.25(=$R_0$)  & 6.25(=$R_0$)   & 6.25(=$R_\mathrm{e0}$)  & 1.0(=$R_\mathrm{e0}$)   \\
Number density (m$^{-3}$) & $1.0\times 10^{17}$ & $1.0\times 10^{13}$ & $1.0\times 10^{13}$  & $3.2\times 10^{14}$ \\
Bulk velocity (m/s)       & 200   & 30,000 & 0     & 0     \\
Temperature (K)           & 300   & 232    & 23210 & 23210 \\
Total current (mA)        & -     & 0.59(=$I_\mathrm{i0}$)  & 4.65(=$I_\mathrm{e0}$)  & 4.65(=$I_\mathrm{e0}$)  \\
\hline \hline
\multicolumn{5}{l}{$^a$ The source centers are ($x_0$, $y_0$, $z_0$) for Xe and Xe$^+$ and ($x_\mathrm{e0}$, $y_\mathrm{e0}$, $z_\mathrm{e0}$) for e$^-$.} \\
\end{tabular}
\label{tab:thrustercondition}
\end{table}

In this study, the following assumptions are made to simplify the model. Since the holes on the thruster optics are not modeled, ions are uniformly emitted from the thruster wall surface. In reality, the holes cause variations in ion density, and electrons present to some extent inside the hole~\cite{Perales-Diaz2021-ci, Miyasaka2012-cy}. Unlike a hollow cathode, electrons are emitted from the neutralizer wall surface similar to a filament neutralizer~\cite{Polansky2013-nv, Conde2022-zp}, a direct emission neutralizer~\cite{Wirz2003-gd}, or a diode mode neutralizer~\cite{Kottke2019-qa}. Finally, collisions between Xe and e$^{-}$, Coulomb collisions between Xe$^{+}$ and e$^{-}$, the presence of multiply-charged ions (i.e., Xe$^{++}$ or more), and electron emission due to secondary electron emission (SEE) and ion-induced electron emission (IIEE) are neglected.

\subsection{Case Descriptions}
\label{subsec:casedescription}

We test eight conditions to investigate the effects of in-space versus in-chamber geometries, neutralizer locations, and the presence of background neutral particles, as defined in Table~\ref{tab:caseid}. The in-space geometry simulation, designated as "1," is compared to the in-chamber geometry simulation, designated as "2," for the neutralizer types A and B. The difference in the BC settings was described in Section~\ref{subsec:bcsetting}. Then we discuss the ion beam and neutralizer coupling, changing the neutralizer exit positions at four locations: 2B-0, 2B-1/4, and 2B-1/2 cases indicating $\frac{d_0}{R_0}=0$, $\frac{d_0}{R_0}=\frac{1}{4}$, and $\frac{d_0}{R_0}=\frac{1}{2}$, respectively, where 
 $d_0$ is the distance between the neutralizer exit and thruster exit, and $R_0$ is the thruster radius. The above-mentioned cases do not model neutrals to differentiate the electrical effect from the high-background pressure effect. Therefore, we also simulate a case with a finite background pressure designated as 2B-BP with $\frac{d_0}{R_0}=1$.

\begin{table}[hbt!]
\caption{Test conditions of each case ID for facility effect and neutralizer position study.}
\centering
\small
\begin{tabular}{l|ccc}
\hline \hline
Case ID$^*$ & Outer boundary & Distance, $d_0/R_0$ & Background neutral  \\  \hline
1A      & In-space       & 0               & Not present   \\
1B      & In-space       & 1               & Not present   \\
2A      & In-chamber     & 0               & Not present   \\
2B      & In-chamber     & 1               & Not present   \\
2B-0    & In-chamber     & 0               & Not present   \\
2B-1/4  & In-chamber     & 1/4             & Not present   \\
2B-1/2  & In-chamber     & 1/2             & Not present   \\
2B-BP   & In-chamber     & 1               & Present       \\
\hline \hline
\multicolumn{4}{l}{$^*$ The neutralizer types, A and B, are described in Table~\ref{tab:thrustercondition}.} \\
\end{tabular}
\label{tab:caseid}
\end{table}

\begin{figure}[hbt!]
    \centering
    \includegraphics[width=0.5\textwidth]{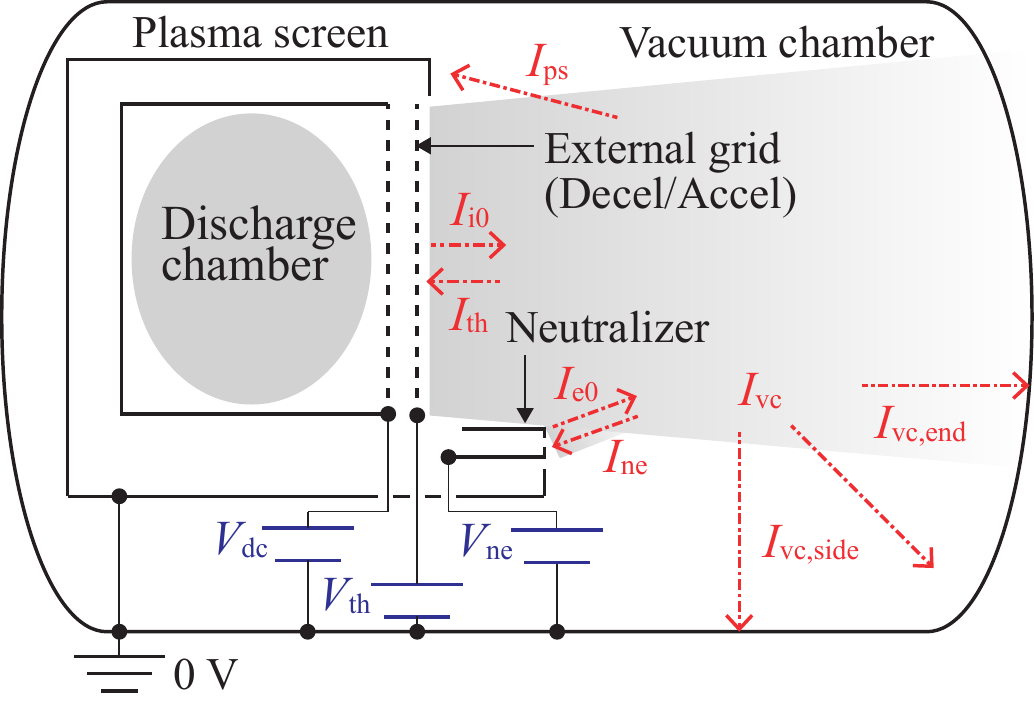}
    \caption{Electrical schematics of this study. $I_\mathrm{i0}$ and $I_\mathrm{e0}$ are the currents of ion emitted from the thruster and electrons emitted from the neutralizer, respectively. $I_\mathrm{ne}$ is the current to the neutralizer exit, $I_\mathrm{th}$, $I_\mathrm{ps}$, and $I_\mathrm{vc}$ are the currents to the thruster exit, plasma screen, and vacuum chamber, respectively. $V_\mathrm{dc}$, $V_\mathrm{th}$, and $V_\mathrm{ne}$ are the biased voltages of the discharge chamber, thruster external gird, and neutralizer exit, respectively.}
    \label{fig:elecschem}
\end{figure}

An electrical schematic diagram for GIT ground operation modeled in this study is shown in Fig.~\ref{fig:elecschem}. Ions with a potential of $V_\mathrm{dc}$ generated in the discharge chamber inside the thruster are accelerated by ion optics and emitted from the external grid with a voltage of $V_\mathrm{th}$, where the beam ion kinetic energy in the axial direction corresponds to $V_\mathrm{dc}-V_\mathrm{th}$. Since this study uses 30,000 m/s for the beam ion velocity when $V_\mathrm{th}=0$~V, $V_\mathrm{dc}$ is set at 612~V, assuming that there is only ion motion outside the thruster, for simplicity. This study further investigates how the thruster plume is affected by using similar $V_\mathrm{dc}$ and $V_\mathrm{th}$ potential conditions to that in actual experiments. Table~\ref{tab:potentialcaseid} shows these three additional conditions with respect to the 2B-BP case. The 2B-ACC case corresponds to the situation where there is no outermost decel grid so that the accel grid with $V_\mathrm{th} = -200$~V is exposed to the plasma plume. The exit density and velocity for the 2B-ACC case are corrected to $8.68\times 10^{12}$ m$^{-3}$ and 34,553 m/s, respectively, because the incoming ions are considered to have a 200~V higher axial energy compared to the baseline case 2B-BP. The 2B-NM and 2B-NP cases simulate the case where the neutralizer is negatively and positively biased relative to the chamber and plasma screen by $V_\mathrm{ne}$, which corresponds to the neutralizer coupling voltage applied in GIT experiments. 

\begin{table}[hbt!]
\caption{Test conditions of each case ID for electric potential boundary study.}
\centering
\small
\begin{tabular}{l|cc}
\hline \hline
\multirow{2}{*}{Case ID} & \multicolumn{2}{c}{Electric potential} \\
 & Thruster exit, $V_\mathrm{th}$ & Neutralizer exit, $V_\mathrm{ne}$  \\  \hline
2B-BP      & 0~V    & 0~V   \\
2B-ACC$^*$ & -200~V & 0~V   \\
2B-NM      & 0~V    & -5~V  \\
2B-NP      & 0~V    & 5~V   \\
\hline \hline
\multicolumn{3}{p{9 cm}}{$^*$The ion bulk velocity is corrected to 34,553 m/s, and the ion exit density to $8.68\times 10^{12}$ m$^{-3}$. All other conditions are the same as the 2B-BP case.} \\
\end{tabular}
\label{tab:potentialcaseid}
\end{table}

After the simulations reach a steady state, the current is calculated for ions and electrons by directly sampling the computational particles, as shown in Fig.~\ref{fig:elecschem}. From the conservation laws, ion and electron currents are given by:
\begin{equation}
    I_\mathrm{i0} = \left\{I_\mathrm{ne}+I_\mathrm{th}+I_\mathrm{ps}+I_\mathrm{vc}\right\}_\mathrm{ion},
\end{equation} 
\begin{equation}
    I_\mathrm{e0} = \left\{I_\mathrm{ne}+I_\mathrm{th}+I_\mathrm{ps}+I_\mathrm{vc}\right\}_\mathrm{electron}. 
\end{equation}

CHAOS has multiple GPUs with MPI-Cuda parallelization strategies~\cite{Jambunathan2018-xf}. This study uses 16 NVIDIA A100 GPUs on the Delta machine at the National Center for Supercomputing Applications for all cases. In the cases without neutral particles, we simulate 500,000 steps before sampling and then sample 200,000 steps to obtain the field macro-parameters and currents. In contrast, in the case with neutral particles, we simulate 5,000,000 steps prior to sampling due to the slow CEX particle motion and then sample 500,000 steps. The total simulation runtimes are; about 30 hours for the 1A and 1B cases, about 12 hours for the 2A, 2B, 2B-0, 2B-1/4, and 2B-1/2 cases, about 90 hours for the 2B-BP, 2B-ACC, 2B-NM, and 2B-NP cases.


\section{Effect of Space vs. Ground-Based Chamber Conditions}
\label{sec:result-spchamber}

\begin{figure}[hbt!]
    \centering
    \includegraphics[width=0.7\textwidth]{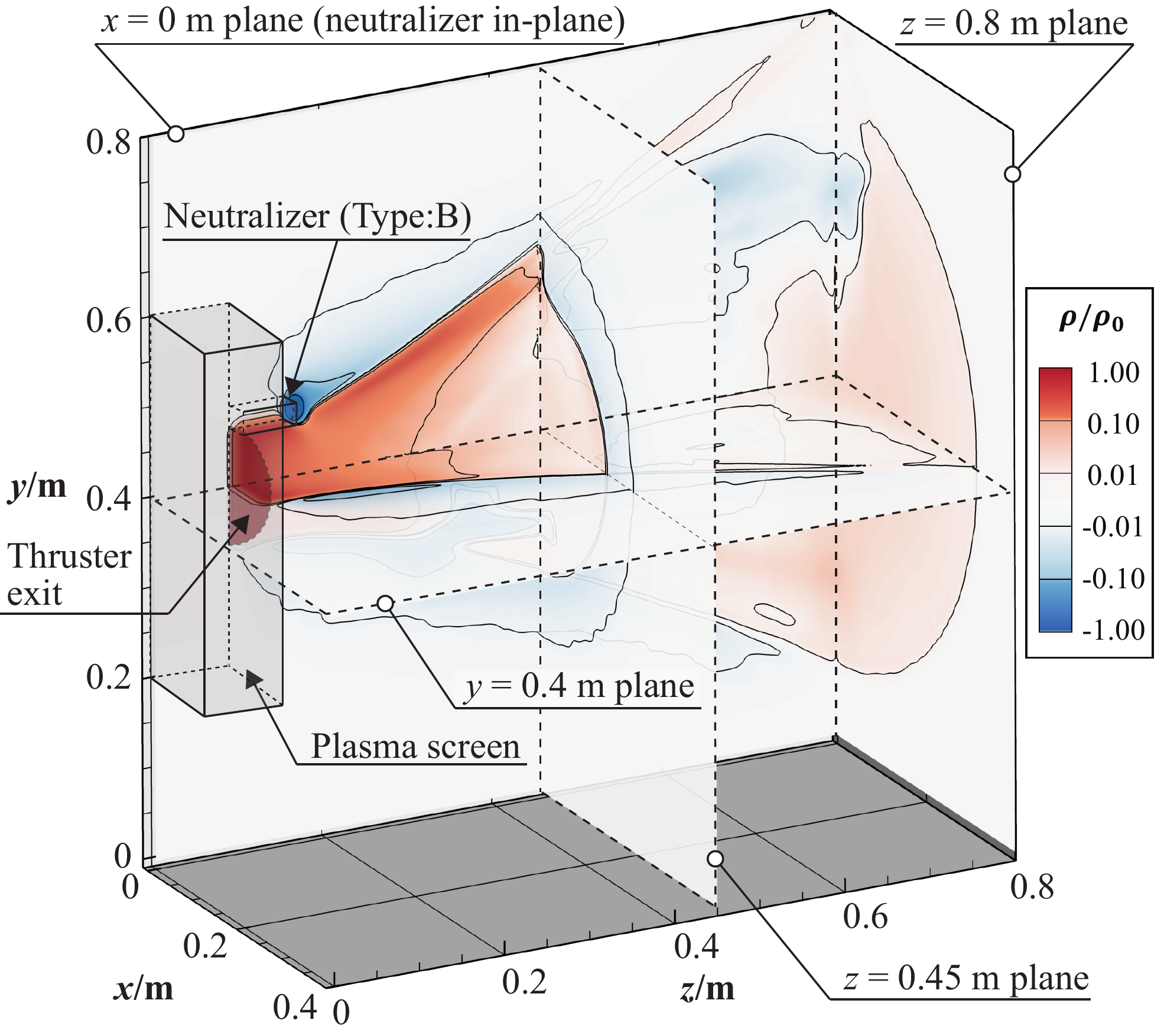}
    \caption{Three-dimensional normalized volume charge density, $\rho/\rho_0$, for the 2B case. (Chamber walls are not shown, for clarity.)}
    \label{fig:3drhodist_2B}
\end{figure}

This section presents the outer edge boundary effect between the in-space and in-chamber cases. Figure~\ref{fig:3drhodist_2B} shows the volume charge density, $\rho$, of the 2B case to highlight the three-dimensional plume structure, where $\rho$ is normalized by the reference charge density $\rho_0=e n_0$. Ion beams with a velocity of 30,000 m/s and thermal electrons with a temperature of 2~eV are emitted from the thruster and neutralizer exit in the positive $z$-direction, resulting in a maximum and minimum $\rho$ near the respective exit points. As the ion beam moves downstream, it is neutralized by coupling with the electrons but also expands due to the positive space charge. In the three-dimensional diagram, we indicated two planes that will be used for comparing subsequent cases.

\begin{figure}[hbt!]
    \centering
    \includegraphics[width=\textwidth]{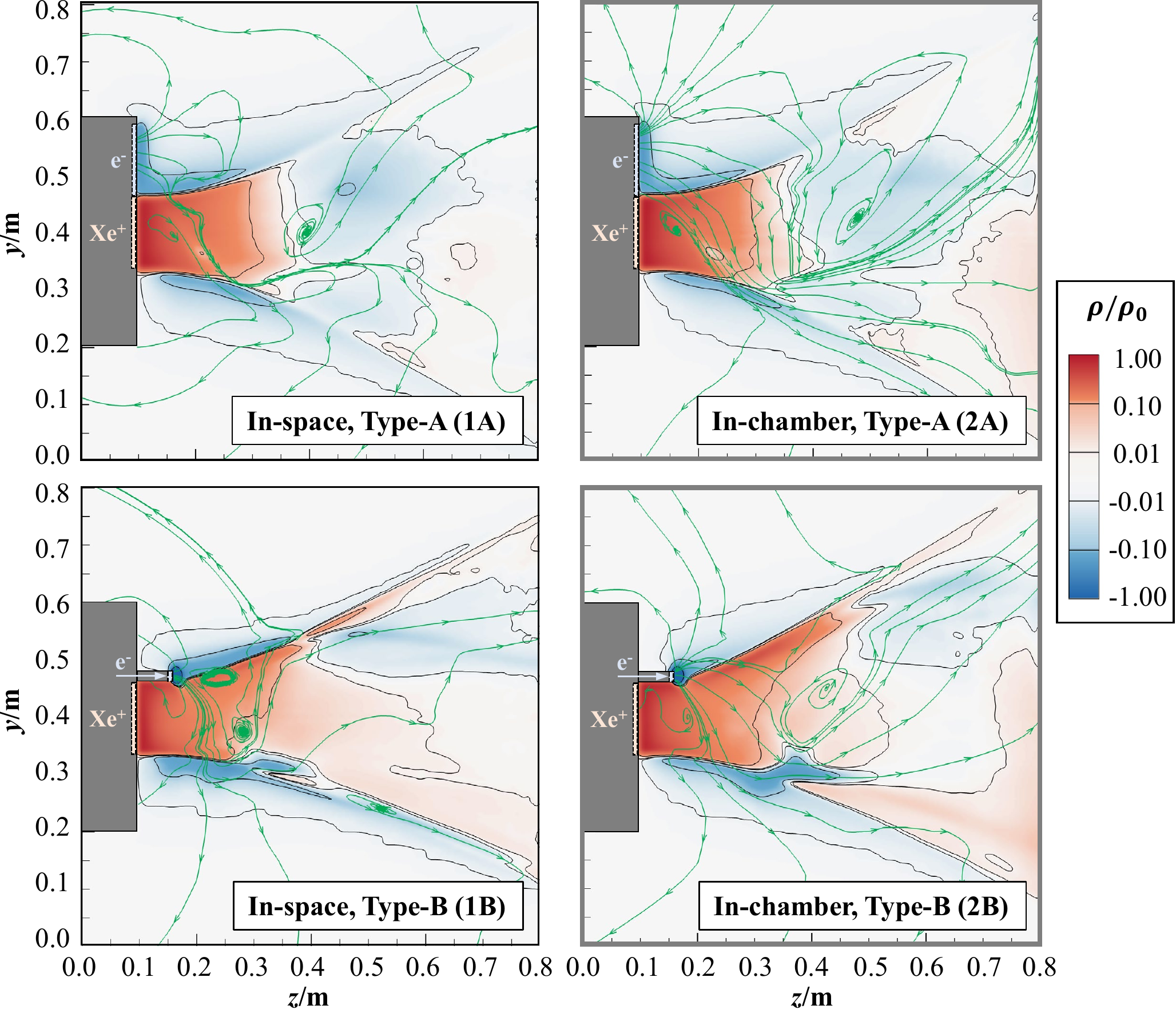}
    \caption{Volume charge density in the $x=0$~m plane for the in-space vs. in-chamber cases for neutralizer types A and B, respectively. The green lines show electron streamlines based on electron $z$- and $y$-velocities.}
    \label{fig:rhodist_spacechamber}
\end{figure}

Not only the type B, as shown in Fig.~\ref{fig:3drhodist_2B}, but also the type A neutralizer are examined with different BCs at the edge of the domain, i.e., 1A versus 2A and 1B versus 2B. Figure~\ref{fig:rhodist_spacechamber} shows the charge density in the $x=0$~m plane, with color contours indicating the degree of neutralization of the plume and green arrows indicating the electron streamlines in the plane. Regardless of the neutralizer type, the in-chamber results show positive charge density near the downstream wall (0.7~m $< z <$ 0.8 m) due to the formation of a sheath by the charge-absorbing wall at 0 V. In addition, the green arrows indicate a higher density of electron streamlines that flow downstream compared to the in-space case. These are an example of facility effects caused by the presence of grounded potential walls. It is also evident that the same type of neutralizer as in the previous study~\cite{Jambunathan2020-yx} (type A) and the more realistic type of neutralizer (type B) create different spatial distributions of the volume charge density both in the in-space and in-chamber cases, even though the electron emission is set to the same level. This suggests that the type and location of the neutralizer significantly affect the neutralization of the plume, as discussed further in Section~\ref{sec:result-neutpos}.

\begin{figure}[hbt!]
    \centering
    \includegraphics[width=\textwidth]{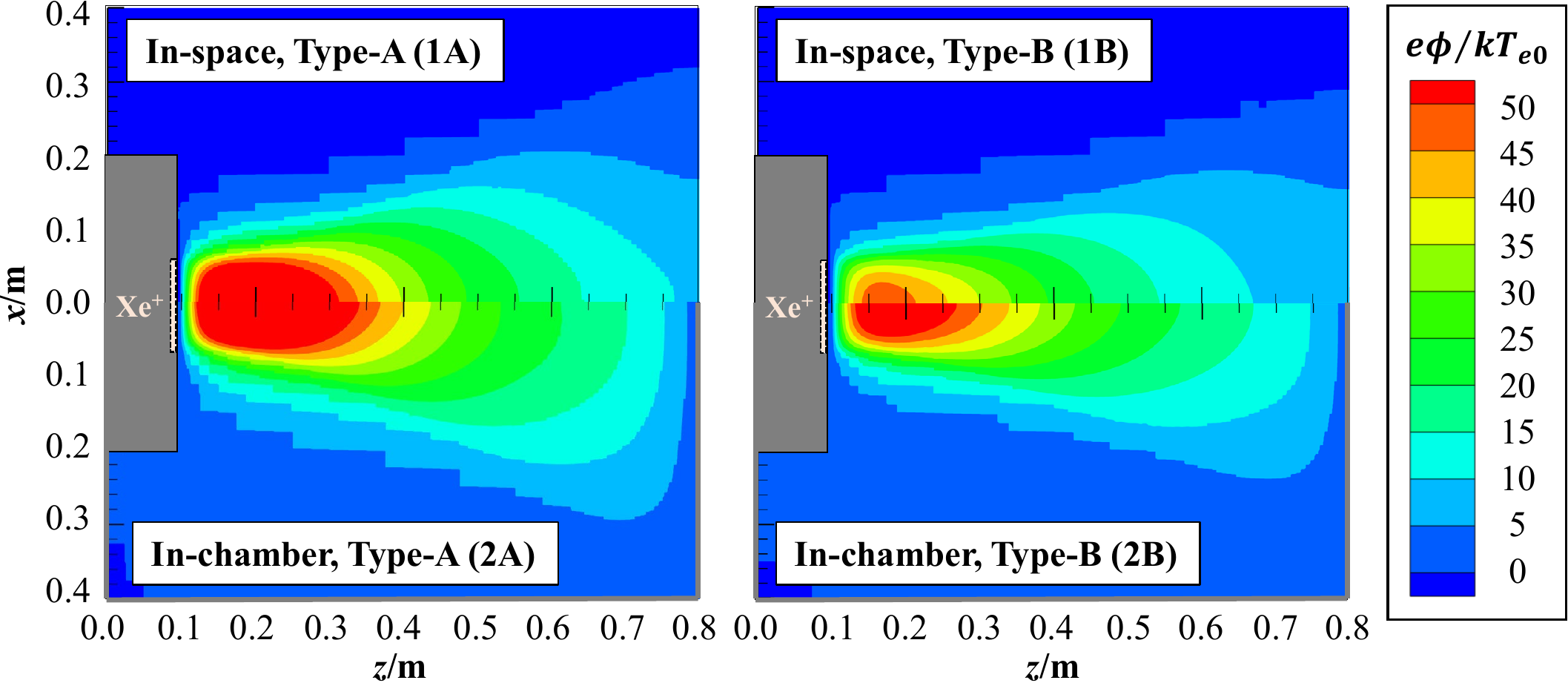}
    \caption{Electric potential in the $y=0.4$~m plane for the in-space vs. in-chamber cases for neutralizer type A and type B, respectively.  Here, and in similar subsequent figures, $kT_{e0} = $2~eV.}
    \label{fig:phi_spacechamber}
\end{figure}

\begin{figure}[hbt!]
    \centering
    \includegraphics[width=0.5\textwidth]{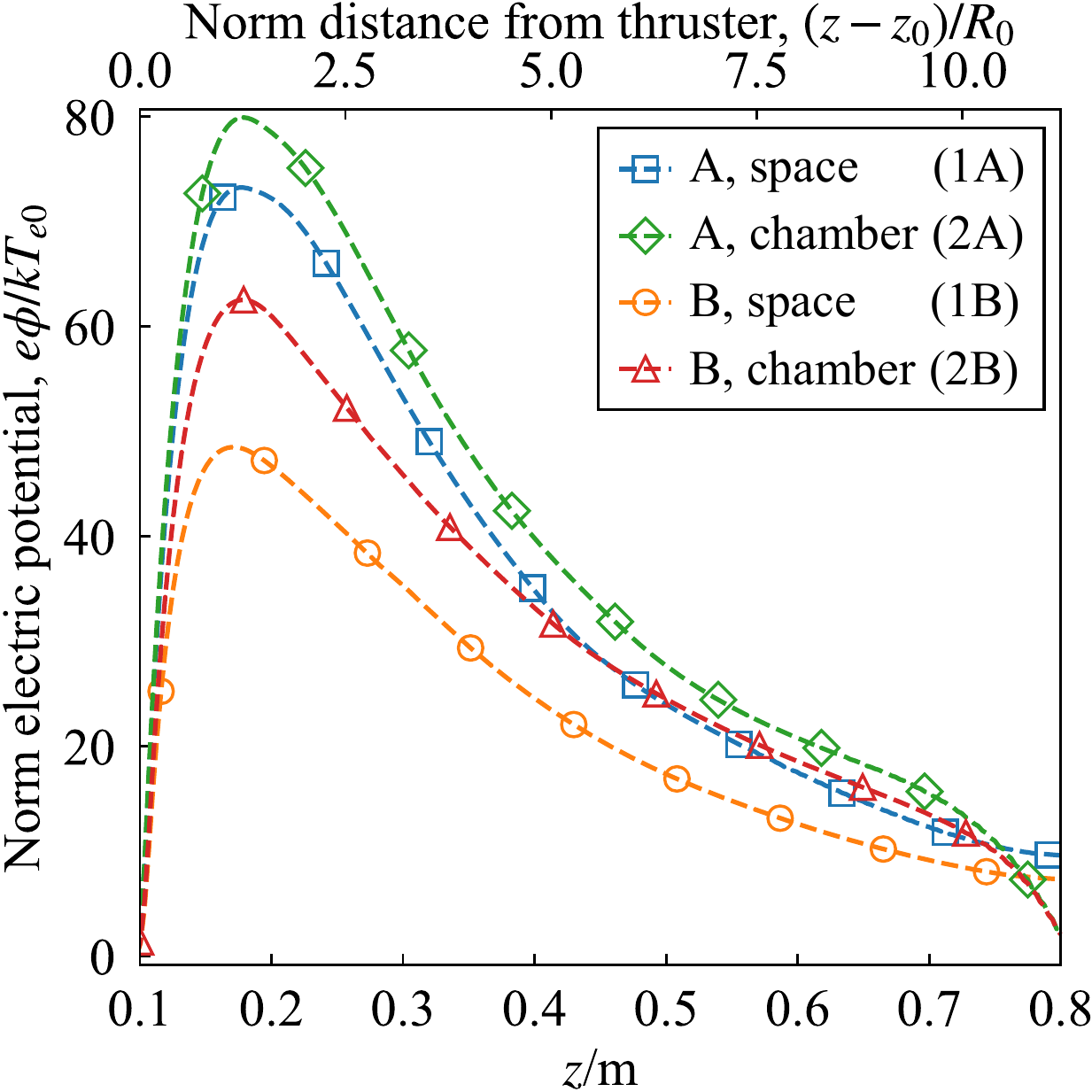}
    \caption{Electric potential along the thruster axis for the in-space vs. in-chamber cases for neutralizer type A and type B, respectively.}
    \label{fig:SpaceCham_phi_thruster}
\end{figure}

Figure~\ref{fig:phi_spacechamber} shows the potential, $\phi$, in the $y=0.4$~m plane, where $\phi$ is normalized by $kT_\mathrm{e0}/e=2$ V. The potential in the $y=0.4$~m plane increases from the thruster exit downstream to a maximum value and then decreases further downstream. When comparing the in-space and in-chamber cases, higher potentials are observed in the chamber case for both neutralizer types. This is also quantitatively shown in the line plot of the potential along the thruster axis as given in Fig~\ref{fig:SpaceCham_phi_thruster}. The difference in the number of electrons present in the computational domain explains the reason for this behavior. Table~\ref{tab:numofcompspecies} displays the number of computational ions and electrons in the entire domain at steady-state. In all cases, the number of electrons is lower than that of ions, but it is more pronounced in the chamber case. This is because, at the edge of the computational domain, the chamber absorbs all electrons, whereas in the space case, the CCE BC simulates an actual space condition in which there is a backflow of electrons that are trapped in a potential well formed by the 
 ion beam.

\begin{table}[hbt!]
\caption{Number of computational particles in the entire computational domain at the steady state.}
\centering
\small
\begin{tabular}{lcccc}
\hline \hline
Species    & 1A   & 2A   & 1B   & 2B   \\ \hline
Xe$^+$ (M) & 18.4 & 18.6 & 17.8 & 18.1 \\
e$-$   (M) & 17.8 & 16.1 & 17.2 & 16.2 \\
\hline \hline
\end{tabular}
\label{tab:numofcompspecies}
\end{table}

\begin{figure}[hbt!]
    \centering
    \includegraphics[width=0.45\textwidth]{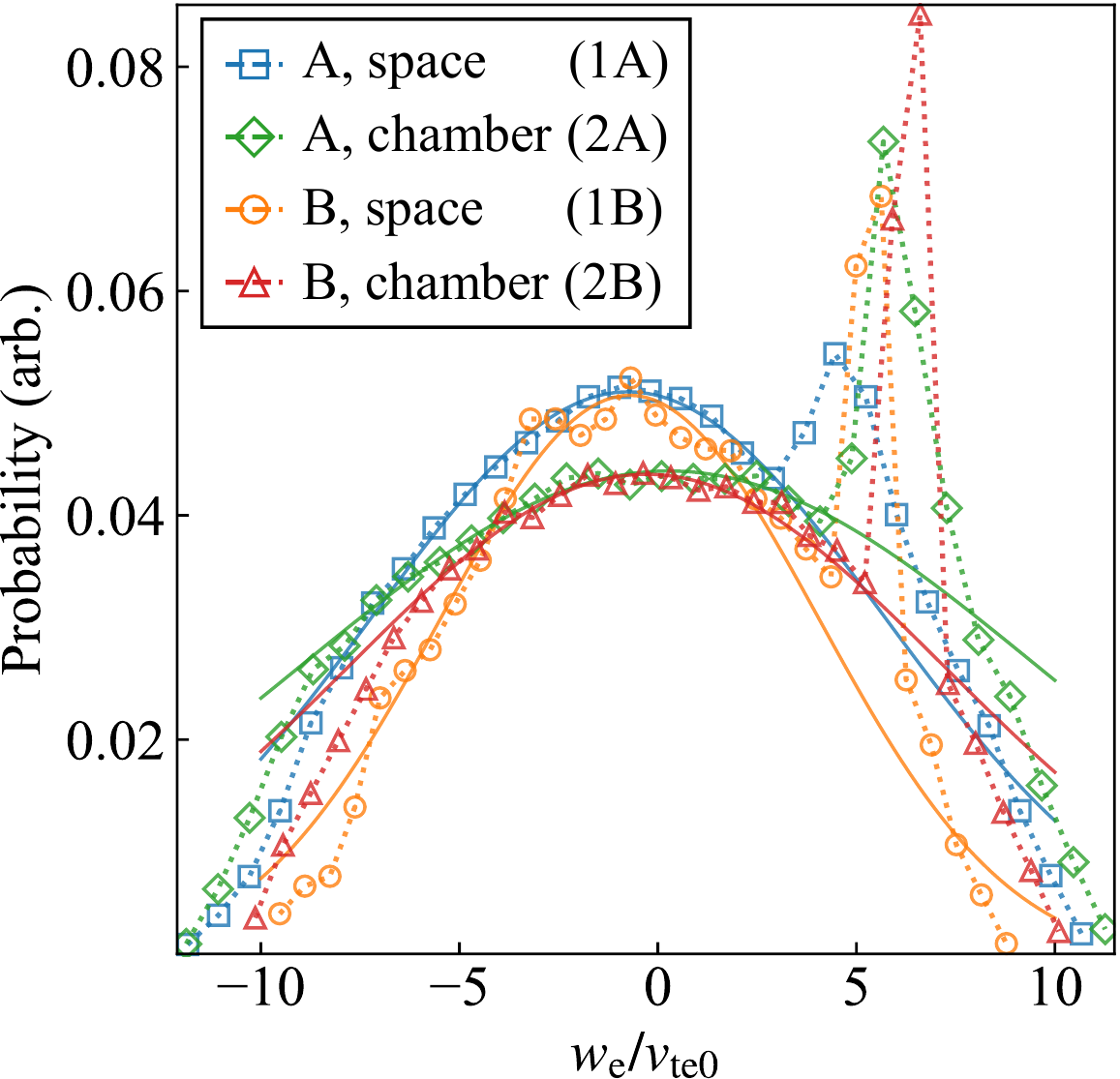}
    \caption{EVDF in $z$-direction at ($x, y, z$) = (0, 0.4, 0.1625)~m for the 1A, 2A, 1B and 2B cases. Computational electrons are sampled at $R_0$ downstream from the thruster exit center. The solid lines are Maxwellian distribution fitting curves based on Eq.~(\ref{eq:maxwellian}), and the normalized electron temperatures obtained by fitting, $T_\mathrm{e}/T_\mathrm{e0}$, are 21~eV (1A), 11~eV (1B), 43~eV (2A), and 28~eV (2B), where $T_\mathrm{e0}=2$ eV.}
    \label{fig:EDF_spacechamber}
\end{figure}

Since the behavior of electrons is key to understanding the difference in the spatial variation of $\phi$, it is important to evaluate the electron velocity distribution functions (EVDFs) as well as the macro parameters of the electron flow indicated by the green arrows in Fig.~\ref{fig:rhodist_spacechamber}. Figure~\ref{fig:EDF_spacechamber} shows the $z$-direction EVDFs obtained by sampling computational electrons at ($x, y, z$) = (0, 0.4, 0.1625)~m, i.e., at a position $R_0$ downstream along the thruster axis. The dotted lines and markers indicate the fraction of electrons in a velocity bin, where the velocity and temperature values are normalized by the reference electron thermal velocity, $v_\mathrm{te0}$, of 838,782 m/s, and the electron temperature at the neutralizer exit, $T_\mathrm{e0}$, of 23,200 K (2 eV), respectively. In all cases, there are clearly two distributions for the electrons. One is a thermalized distribution with a high temperature peaking at nearly $w_\mathrm{e}/v_\mathrm{te0}=0$, and the other is a cold electron flow with a peak around $w_\mathrm{e}/v_\mathrm{te0}$ = 5. This flow can be understood from the streamlines moving toward the lower right at ($z$, $y$) = (0.2, 0.4)~m in Fig.~\ref{fig:rhodist_spacechamber}. In other words, there are two types of electrons in the steady-state ion thruster plume: thermal electrons trapped around the plume core and electrons flowing downstream driven by the potential gradient. 

The solid lines in Fig.~\ref{fig:EDF_spacechamber} are fitting curves obtained by a one-dimensional Maxwellian distribution function of:
\begin{equation}
    \label{eq:maxwellian}
    f(w_\mathrm{e})= \left(\frac{m_\mathrm{e}}{2\pi k T_\mathrm{e}}\right)^{1/2} e^{-\frac{m_\mathrm{e}w_\mathrm{e}^2}{2 k T_\mathrm{e}}},
\end{equation}
where $w_\mathrm{e}$ is the electron $z$-velocity, $m_\mathrm{e}$ is electron mass, $k$ is Boltzmann's constant, and $T_\mathrm{e}$ is the electron temperature. The fitting range is -6 $< w_\mathrm{e}/v_\mathrm{te0} <$3 in the $z$-direction since only bulk electrons are considered. The obtained normalized fitting temperatures $T_\mathrm{e}/T_\mathrm{e0}$ are of the order of about 10 and are larger for the in-chamber cases for both types of neutralizers. This is due to the difference in the maximum potential and the sheath at the chamber downstream wall of the plume creating a steep potential gradient that attracts and accelerates more electrons, as shown in Fig~\ref{fig:SpaceCham_phi_thruster}.

The obtained potentials and electron temperatures are a few times larger than those typically obtained by experiments~\cite{Kerslake1993-my, Polansky2013-nv, Conde2022-zp, Yizhou2017-ee} because some actual geometries and physics are neglected in the model of this study as described in the last paragraph of Section~\ref{subsec:geomandspecies}. First, SEE may be an important factor, especially on the chamber downstream wall at $z=0.8$~m, where electrons incident with an energy of 6.37~eV on average are observed in this study. In this energy region, according to Ref.~\cite{Bellissimo2020-qf}, the total emission yield from carbon is between 0.1--0.5, which means more than 10\% of electrons are recovered when they hit the wall, 
although this estimate requires a large extrapolation of their data to the much lower energies of our case. The small energy secondary electrons emitted from the grounded walls would be trapped inside the high potential plume if they are born with energies of 2~eV, another unknown and could reduce the plume potential. Second, since an actual GIT has holes on the exterior grid, neutralization is initiated closer to or inside the exterior grid by electrons inside the grid holes. The third possibility is the axial location of the neutralizer exit. According to Ref.~\cite{Yizhou2017-ee}, the angle between the normal beam axis and the neutralizer axis can affect the plume potential. Finally, another past study~\cite{Ward1968-dc} suggested that the neutralizer-ion beam coupling was enhanced by ions generated in electron-neutral collisions, which is also not modeled in this study.
 

\section{Effect of Neutralizer Position and Background Pressure in In-chamber Simulations}
\subsection{Neutralizer Position}
\label{sec:result-neutpos}

\begin{figure}[hbt!]
    \centering
    \includegraphics[width=\textwidth]{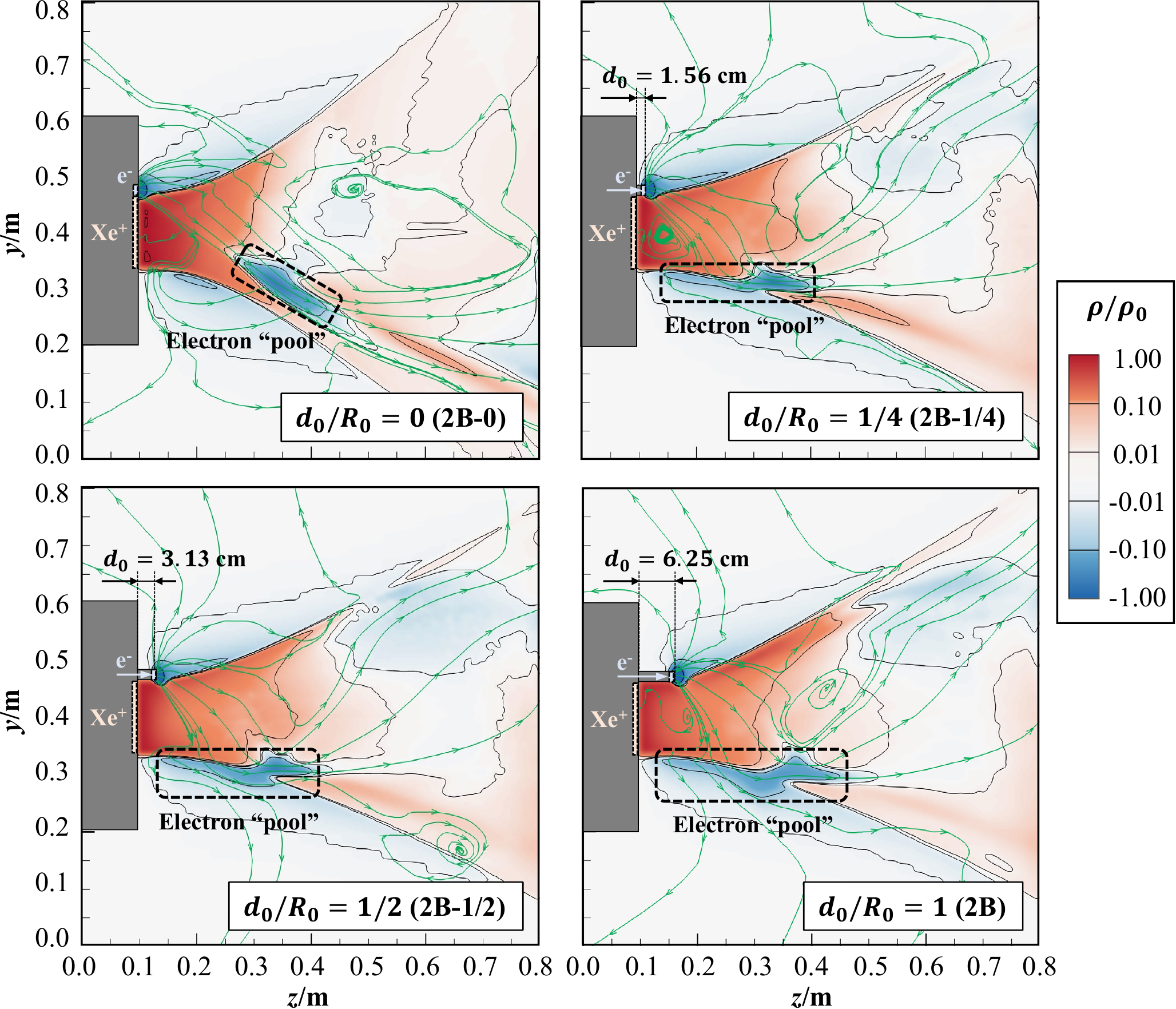}
    \caption{Effect of the location of the type B neutralizer in the vacuum chamber on volume charge density in the $x=0$~m plane. The green lines show electron streamlines based on electron $z$- and $y$-velocities.}
    \label{fig:rhodist_neutpos}
\end{figure}

This section examines how the neutralizer position affects the coupling between ion and electron sources in ground-based testing. Specifically, we consider only the type B neutralizer and perform calculations for the in-chamber case, where the neutralizer position is the only variable, i.e., the cases 2B-0, 2B-1/4, and 2B-1/2. The charge density in the $x=0$~m plane is shown in Fig.~\ref{fig:rhodist_neutpos} for different distances between the neutralizer exit and the plasma screen wall, $d_0$. In all cases, electrons enter slightly above the ion source, form a negative charge density area, and then move toward the lower-right direction, as shown by the green arrows. Electrons begin to slow down and accumulate when they pass through ($y$, $z$) = (0.4, 0.2)~m, where the potential is at its maximum. A negative charge density area around ($y$, $z$) = (0.3, 0.35)~m is formed on the opposite side of the electron source from the point of maximum potential, which we refer to as an ``electron pool'' in this study. This electron pool has also been observed in previous studies~\cite{Usui2013-al, Brieda2005-kj} and is a unique phenomenon in ion thruster plume neutralization with neutralizers adjacent to the thruster.

\begin{figure}[hbt!]
    \centering
    \includegraphics[width=\textwidth]{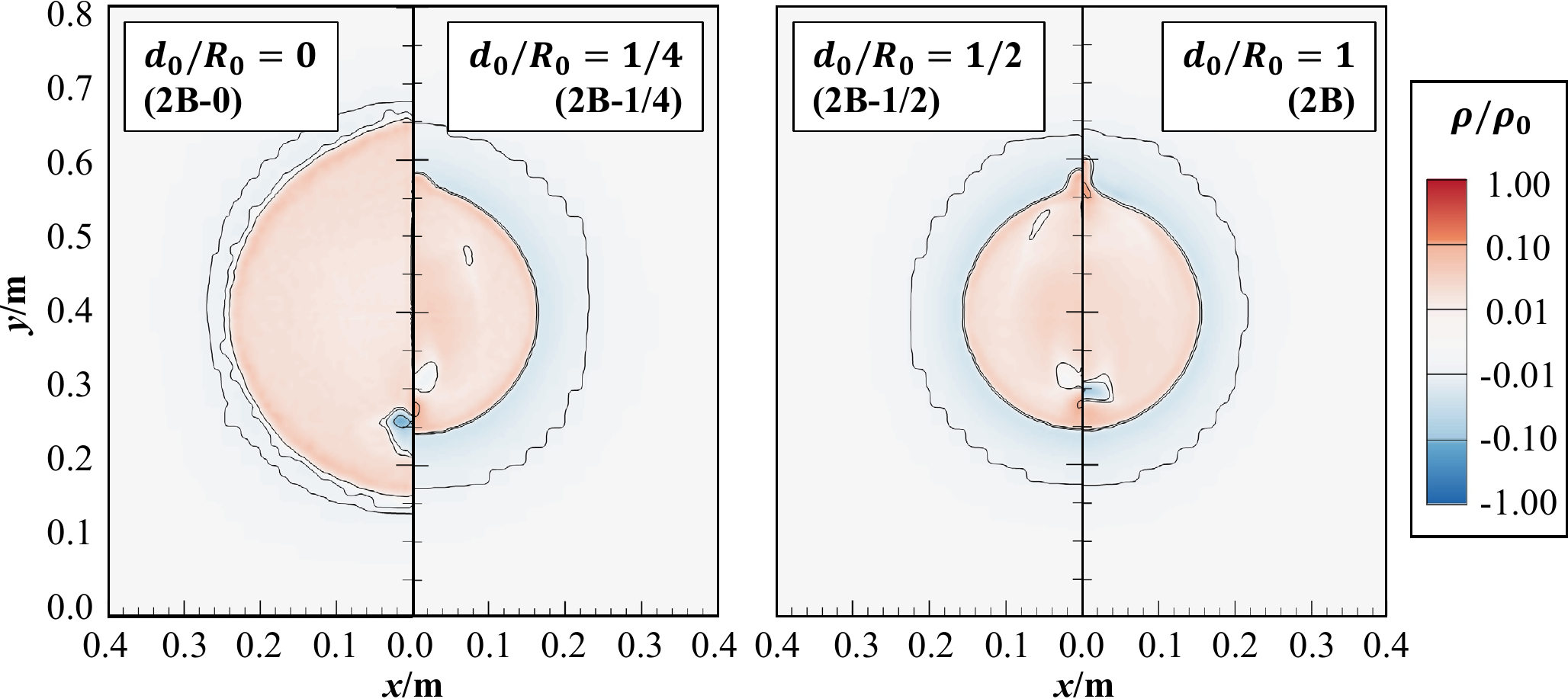}
    \caption{Volume charge density in the $x$-$y$ plane for different $d_0/R_0$ values of the type B neutralizer in the vacuum chamber in the $z=0.45$~m plane.}
    \label{fig:rhodist_neutposxy}
\end{figure}

When $d_0/R_0 = 0$, the electron pool is formed near the plume center, but as $d_0$ increases, the negative region shifts towards the lower right. However, these characteristics of electron motion are mainly observed on the thruster center plane, including the neutralizer exit (the $x = 0$~m plane). Figure~\ref{fig:rhodist_neutposxy} shows the charge density distribution in the $x$-$y$ plane at the $z = 0.45$ m. Although there is significant variation in charge density around $(x,y) = (0,0.25)$~m or $(0,0.6)$~m in Fig.~\ref{fig:rhodist_neutposxy}, the plume is uniformly distributed in most radial directions.

\begin{figure}[hbt!]
    \begin{subfigure}{0.5\textwidth}
        \centering
        \includegraphics[width=0.9\linewidth]{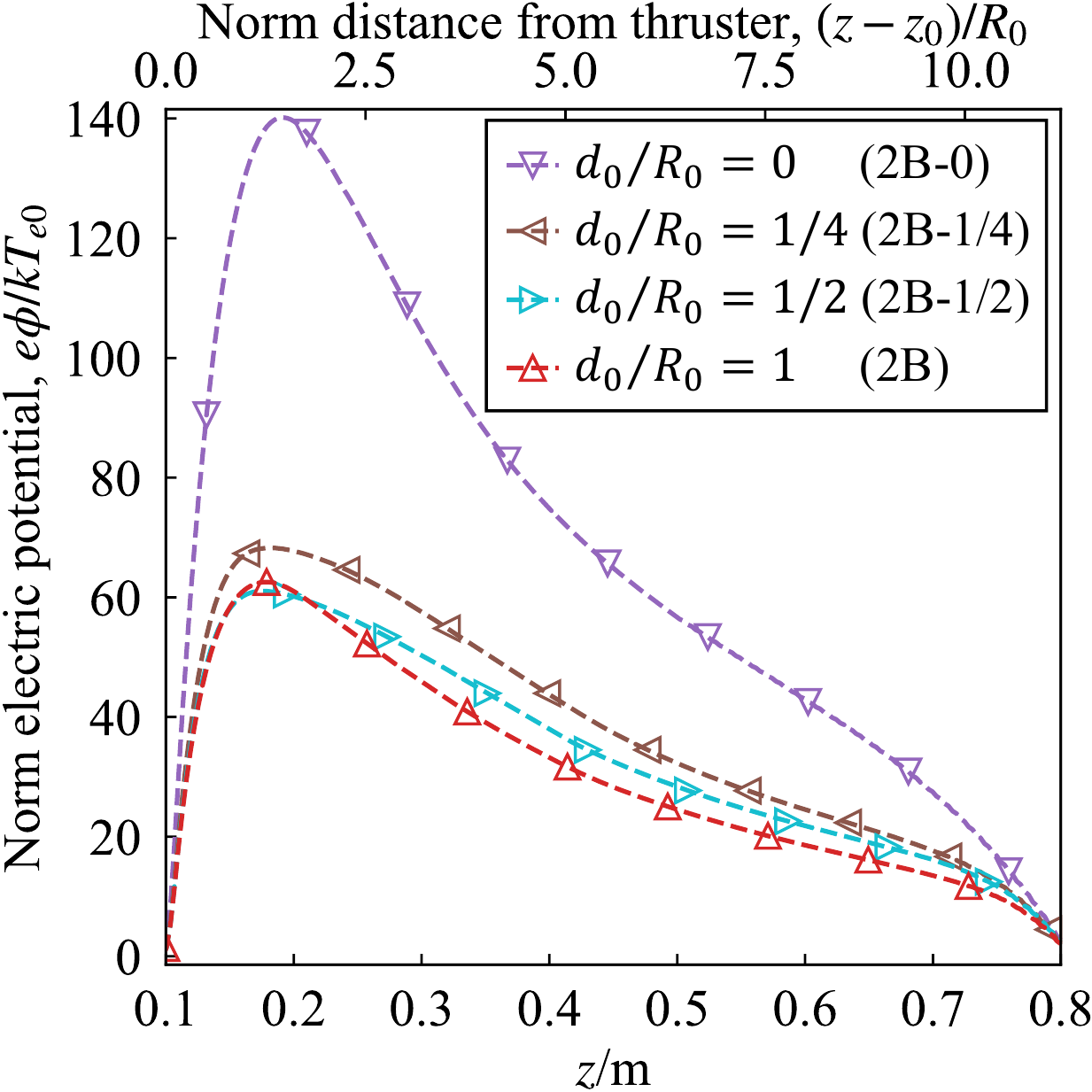}
        \caption{Along the thruster axis, $(x,y)=(0,0.4)$ m.} 
        \label{fig:CathodePos_phi_thruster}
    \end{subfigure}%
    \begin{subfigure}{0.5\textwidth}
        \centering
        \includegraphics[width=0.9\linewidth]{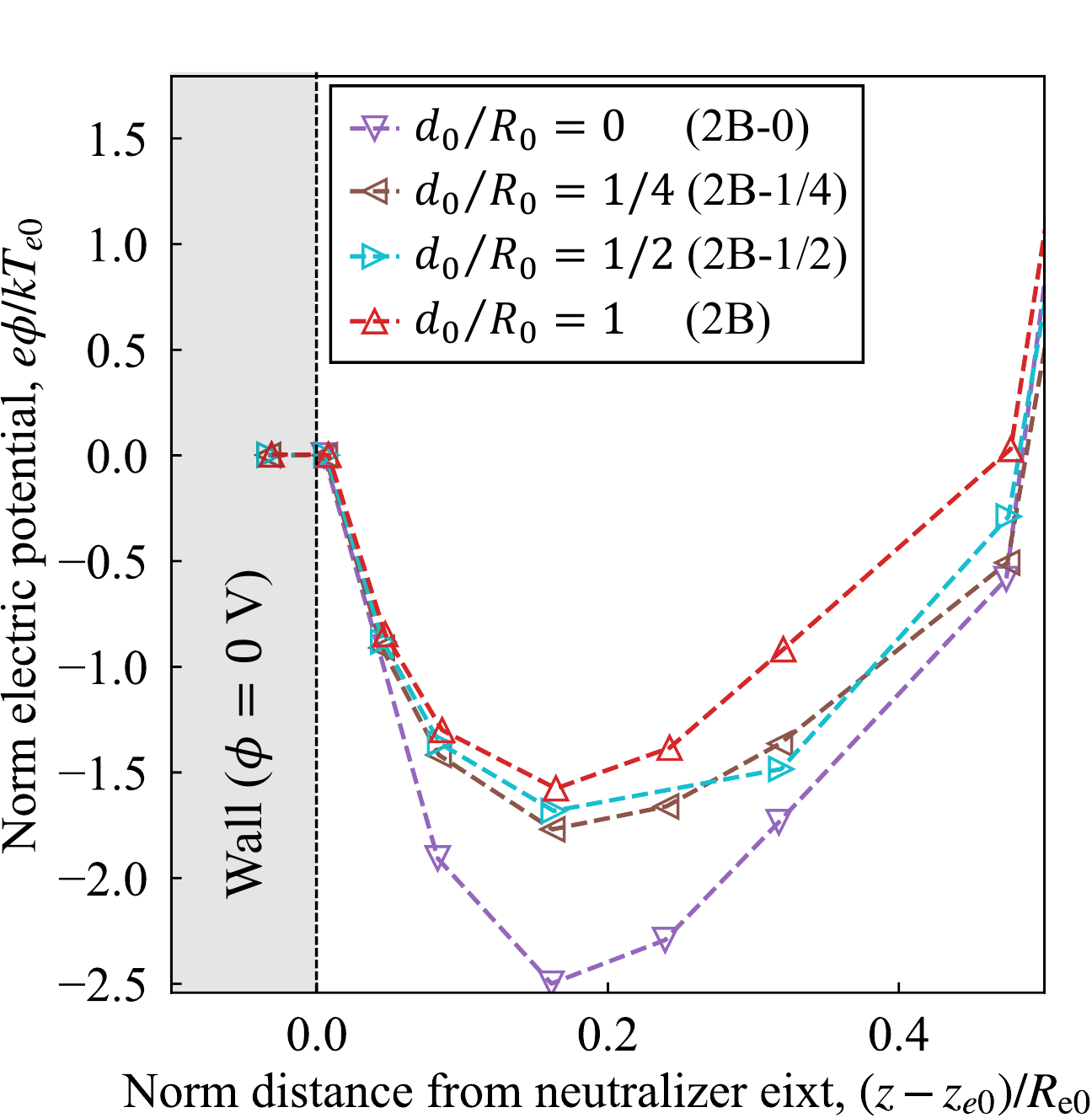}
        \caption{Along the neutralizer axis, $(x,y)=(0,0.473)$ m.} 
        \label{fig:CathodePos_phi_cathode}
    \end{subfigure}%
    \caption{Electric potential along the $z$-direction for different neutralizer locations for the neutralizer Type: B in the vacuum chamber.}
    \label{fig:CathodePos_potentialline}
\end{figure}

Next, Fig.~\ref{fig:CathodePos_phi_thruster} shows the normalized potential along the thruster axis for each neutralizer-type position. 
The plume potential increases as $d_0$ decreases, and the peak potential is particularly high for $d_0/R_0 = 0$. This result indicates that the best neutralizer-ion beam coupling is obtained where $1/2 < d_0/R_0 < 1$ because the position of the maximum potential is seen to occur approximately $(z-z_0)=R_0$, and there is no significant difference between the $d_0/R_0=1/2$ and 1 cases. Figure~\ref{fig:CathodePos_phi_cathode} shows the normalized potential on the neutralizer axis near the neutralizer exit. In all cases, the potential decreases just after the neutralizer exit and increases after the potential reaches a minimum value at approximately $(z-z_\mathrm{e0})/R_\mathrm{e}=0.2$, forming a virtual cathode. The electron space charge limits the low-energy electron transport in the virtual cathode region, as explained in a previous experimental study~\cite{Yizhou2017-ee}. As $d_0$ increases, the minimum potential in Fig.~\ref{fig:CathodePos_phi_cathode} increases. This is due to the relaxation of the space-charge limitation as the exit of the neutralizer approaches the high potential space. In addition, when the neutralizer exit is on the wall ($d_0/R_0 = 0$), there is no path for electrons to travel upstream to the neutralizer exit and into the plume, thereby reducing the neutralizer-ion beam coupling and increasing the electric potential in the plume. The results shown in this section conclude that moving the neutralizer downstream contributes more to neutralizing the ion beam. This trend is consistent with an experiment in which the coupling voltage decreased as the neutralizer moved toward the downstream~\cite{Yizhou2017-ee}.

\subsection{Background Pressure}
\label{sec:result-CEX}

\begin{figure}[hbt!]
    \centering
    \includegraphics[width=0.5\textwidth]{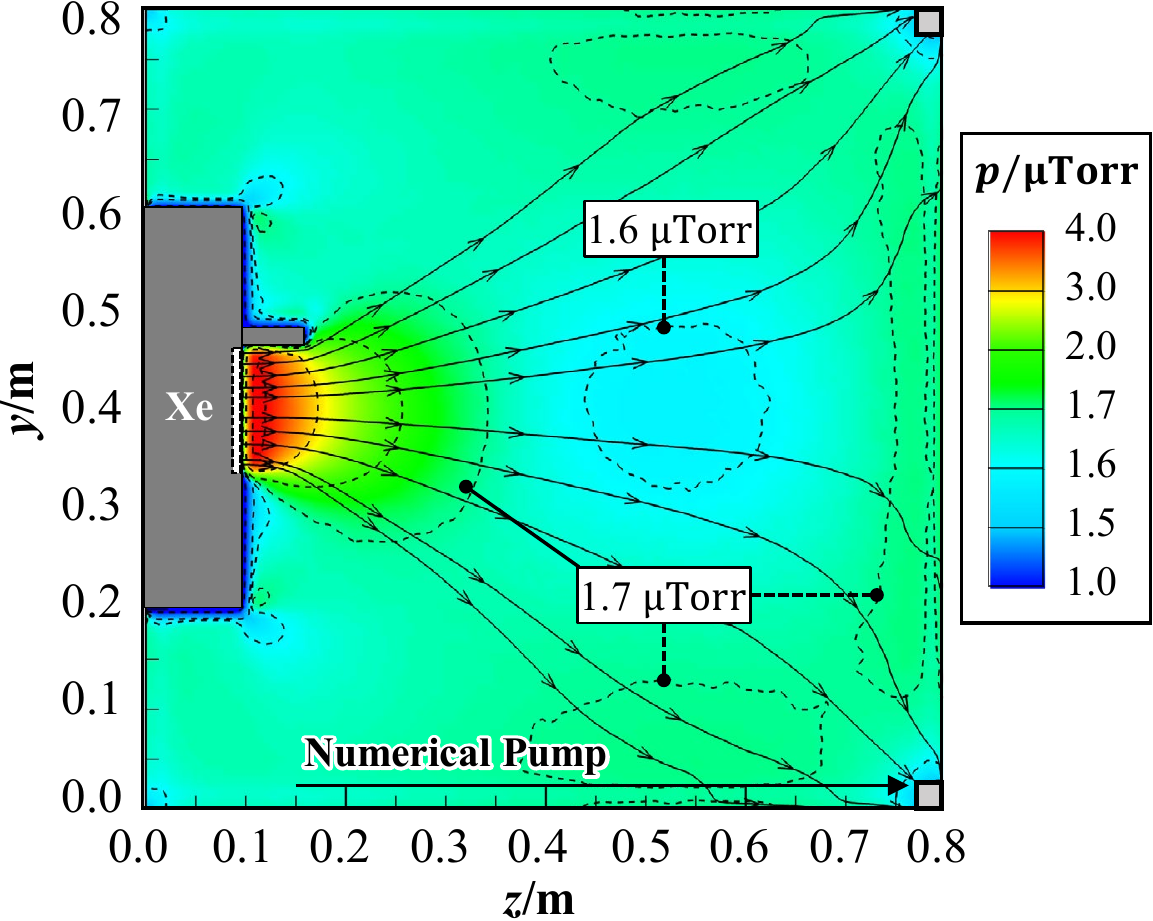}
    \caption{Background pressure distribution for the 2B-BP case in the $x = 0$~m plane.}
    \label{fig:backgroundpress}
\end{figure}

To understand the effect of background pressure in the vacuum chamber, we use the 2B case as the baseline. Figure~\ref{fig:backgroundpress} shows the background pressure, $p$, in the $x=0$~m plane, where $p$, is calculated from the ideal gas equation of $p=n_\mathrm{n}kT_\mathrm{n}$, and $n_\mathrm{n}$ and $T_\mathrm{n}=300$~K are the neutral number density and neutral particle temperature, respectively. As a result of the neutral number density at the thruster exit and the size of the numerical pump, the highest pressure is about 4 \textmu Torr at the thruster exit, with a minimum pressure of about 1.6 \textmu Torr around the center of the vacuum chamber. This is similar to the typical background pressures found during ground test experiments.

\begin{figure}[hbt!]
    \centering
    \includegraphics[width=0.5\textwidth]{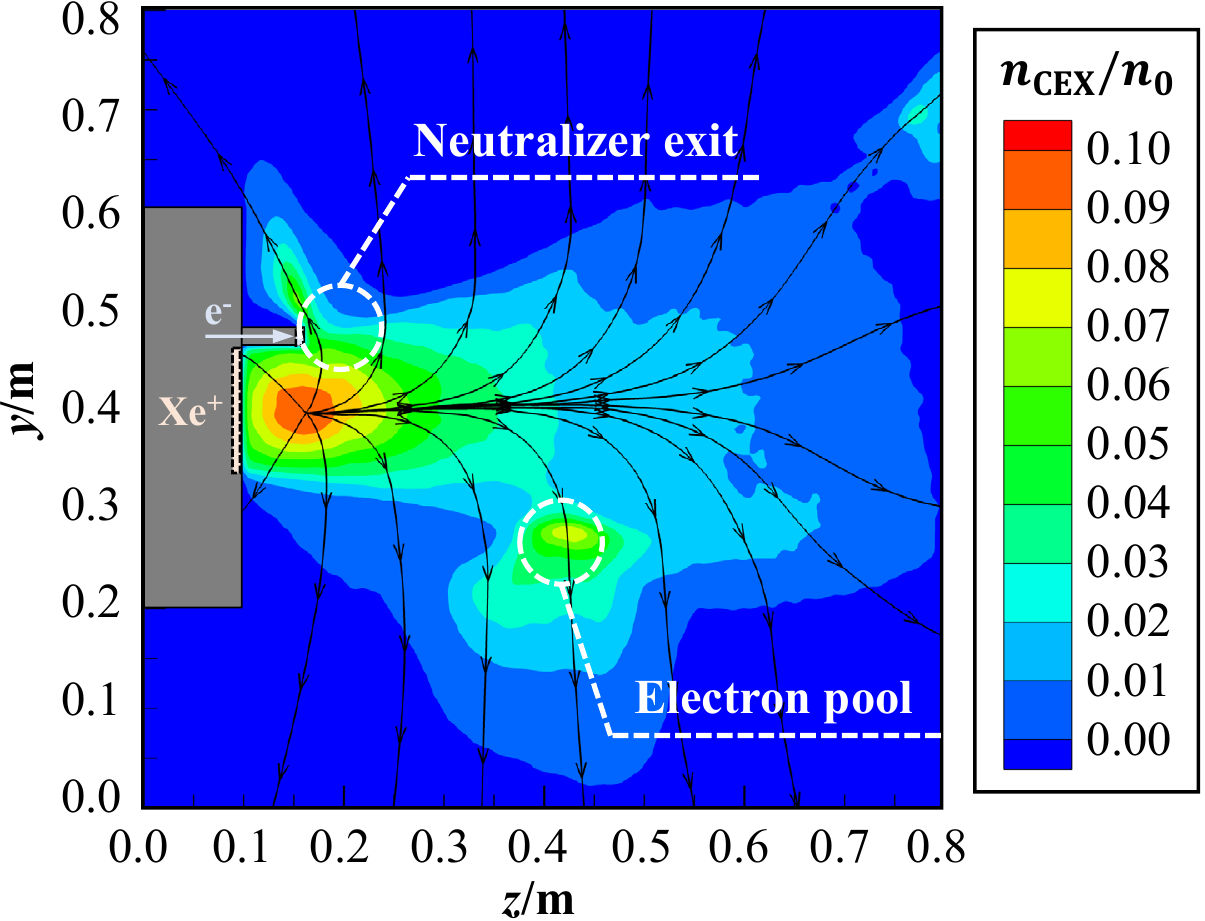}
    \caption{Number density of CEX ions for the 2B-BP case in the $x=0$~m plane.}
    \label{fig:CEXdensity}
\end{figure}

CEX and MEX collisions with the background neutrals change the ion velocity, which, in turn, alters the plume. CEX ions, originally neutral particles, have much smaller velocities than the beam ions and are the main cause of facility effects due to the high-background pressure. Figure~\ref{fig:CEXdensity} shows the number density distribution of CEX ions in the $x=0$~m plane. CEX ions are produced in all regions where beam ions exist, but many CEX ions are particularly produced immediately downstream of the thruster, where the neutral number density is a maximum. However, high-density areas also appear outside the plume core, indicating an asymmetric structure. This occurs because the CEX ions produced with a very small velocity remain in the electron pool region ($y$, $z$) = (0.3, 0.35)~m, as indicated in Fig.~\ref{fig:rhodist_neutpos}. Similarly, CEX ions are trapped in the virtual cathode near the neutralizer exit.

\begin{figure}[hbt!]
    \begin{subfigure}{0.5\textwidth}
        \centering
        \includegraphics[width=0.9\linewidth]{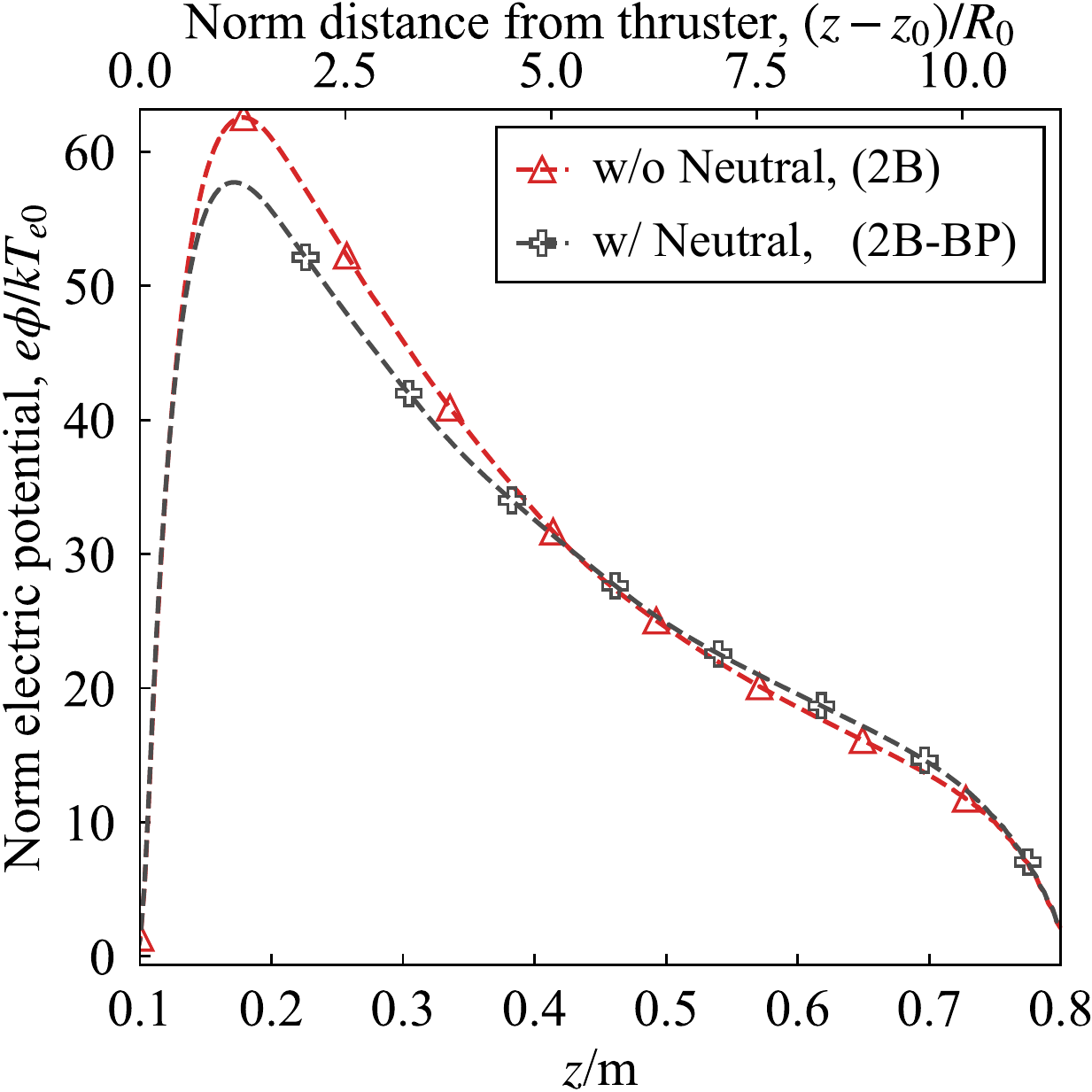}
        \caption{Normalized electric potential} 
        \label{fig:CEX_phi_thruster}
    \end{subfigure}%
    \begin{subfigure}{0.5\textwidth}
        \centering
        \includegraphics[width=\linewidth]{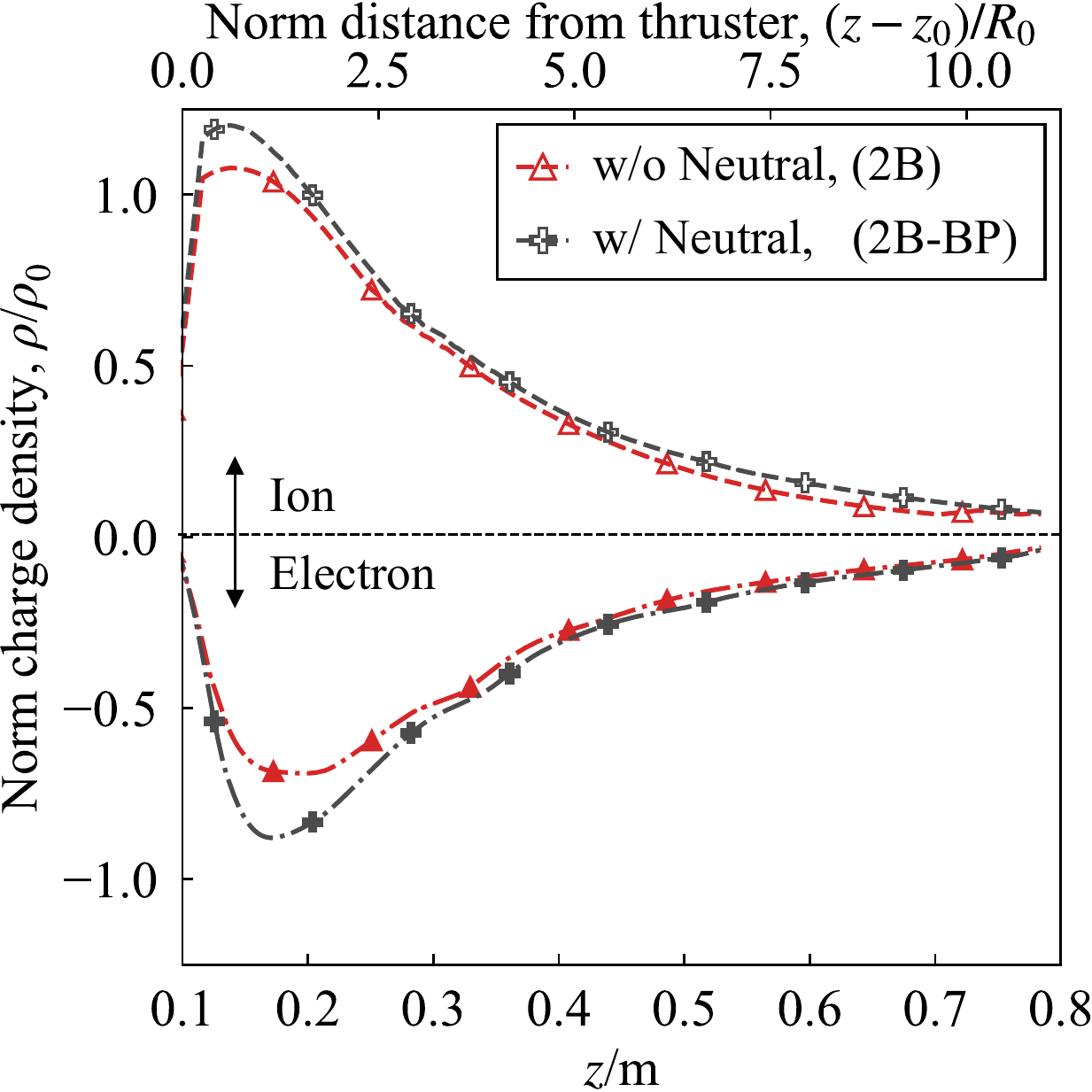}
        \caption{Normalized volume charge density} 
        \label{fig:CEX_rho_thruster}
    \end{subfigure}%
    \caption{Electric potential and volume charge density along the thruster axis for w/ vs. w/o background neutral for the type B neutralizer in the vacuum chamber.}
    \label{fig:CEX_thrustercenterline}
\end{figure}

Figure~\ref{fig:CEX_phi_thruster} shows the potential on the thruster axis, with the maximum potential decreasing in the presence of background neutrals. The decrease is due to two reasons. First, more electrons are present in the plume core. When collisions with neutral particles are modeled, the charge density on the thruster axis for ions and electrons in Fig.~\ref{fig:CEX_rho_thruster} indicates that the presence of CEX ion increases the ion density of the plume by up to 14\%, while the electron density increased by nearly 35\%. The second reason for the decrease in electric potential is the decrease in electron temperature. The electron temperature at ($x, y, z$) = (0, 0.4, 0.1625)~m in the 2B-BP case obtained by a Maxwellian fitting (Eq.~(\ref{eq:maxwellian})) of 15~eV is nearly half that obtained in the 2B case due to the increase in the minimum potential of the virtual cathode. The simulations show that the accumulation of CEX ions in the virtual cathode increases its minimum potential of the virtual cathode from $\phi = -3.2$ to $-2.7$~V. The smaller decrease in voltage at the virtual cathode means that fewer electrons return to the neutralizer, and the kinetic energy of the electrons that can pass through the virtual cathode is lower, resulting in greater neutralization of the plume.

\begin{figure}[hbt!]
    \centering
    \includegraphics[width=0.5\textwidth]{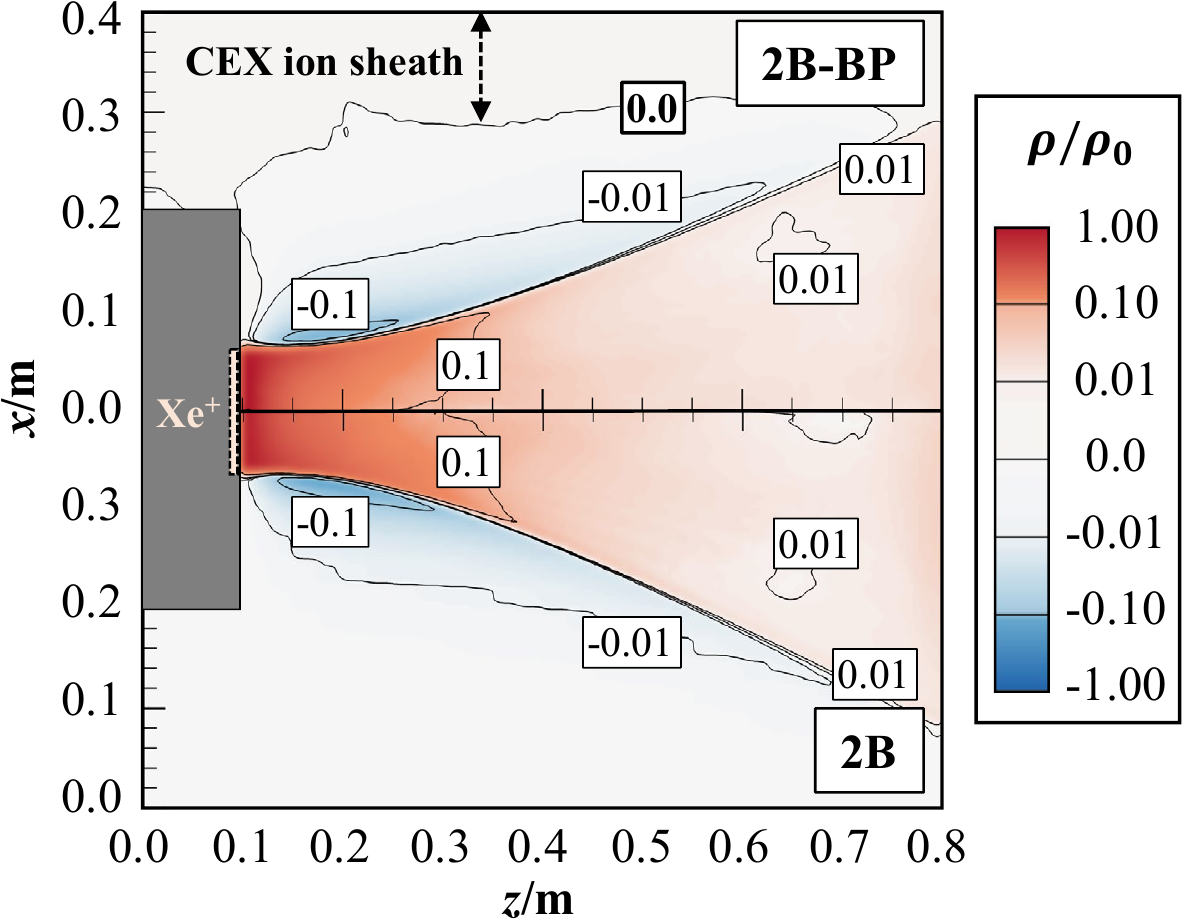}
    \caption{Volume charge density in the $y=0.4$~m plane for a neutral background (2B-BP) vs., vacuum background (2B) cases.}
    \label{fig:rhodist_CEXzx}
\end{figure}

Another interesting difference is the ion sheath formed on the side walls. Figure~\ref{fig:rhodist_CEXzx} shows the normalized volume charge density, $\rho/\rho_0$, of the 2B and 2B-BP cases. Only when background neutral particles are present, the $\rho/\rho_0=0$ line appears between the chamber side wall and the plume core region. This is due to slow CEX ions that create an ion sheath in front of the chamber wall, where we define the ion sheath as the volume near the wall where $\rho/\rho_0>0$. The thickness of the ionic sheath is approximately 0.1 m. A 1-D analytical expression for the Child–Langmuir sheath thickness, $s$, can be calculated as~\cite{Hershkowitz2005-la}
\begin{equation}
    s = \lambda_\mathrm{D,s}\frac{0.79}{\sqrt{\alpha_i}}\left(\frac{e\phi_\mathrm{s}}{T_\mathrm{e,s}}\right)^{3/4}
\end{equation}
where $\alpha_i=0.61$ is the number density factor~\cite{Hershkowitz2005-la}. $\lambda_\mathrm{D,s}$, $\phi_\mathrm{s}$, and $T_\mathrm{e,s}$ are the local Debye length, the electric potential, and the electron temperature at the sheath edge ($\rho/\rho_0=0$ line), respectively. Using ($x, y, z$) = (0.3, 0.4, 0.3)~m as the reference point, we obtain from our simulations: $n_i = n_e \sim 1.0\times10^{11}$~m$^{-3}$, $T_\mathrm{e,s}\sim 2$ eV, and $\phi_\mathrm{s} \sim 7$~V. Thus, the analytical sheath thickness is calculated as $s \sim 0.086$~m, which is close to our simulated sheath thickness.

Having demonstrated that not only background neutrals but also the difference between the space and ground chamber significantly influence electron transport, electric potential, and electron temperature, we next investigate electrical facility effects in terms of the current flow to each location in the simulation domain for cases 1B, 2B, and 2B-BP. Table~\ref{tab:facilityeffectcurrent} displays the currents for ions and electrons based on the current definitions shown in Fig.~\ref{fig:elecschem}. Using the CCE BC (Case-1B), the in-space condition, electrons below threshold energy, $\langle E_\mathrm{e} \rangle$, are reflected at the computational boundary to simulate an infinitely extended region. This suppresses the electrons that flow out of the boundary, resulting in a large decrease in $I_\mathrm{vc}/I_\mathrm{e0}$, for case 1B versus 2B. In contrast, the current flowing into the plasma screen, $I_\mathrm{ps}/I_\mathrm{e0}$, increases significantly by a factor of 8 for the same case comparison. Therefore, ground tests may underestimate the current that flows back to the spacecraft body. 

Comparing the results with and without background neutral particles in Cases 2B vs. 2B-BP, we find that the current flowing to the plasma screen increased for both ions and electrons due to the backflow of CEX ions. Additionally, we demonstrated that the presence of CEX ions increases the potential of the virtual cathode, which results in a decrease in the current flowing back to the neutralizer. The current flow to the downstream and the side walls ($I_\mathrm{vc,end}$ and $I_\mathrm{vc,side}$, respectively) exhibits an interesting behavior due to the presence of the background neutral particles. With respect to ions, 24\% of the ion currents change their direction toward the side wall (see the difference in $I_\mathrm{i,vc,side}$) due to collisions with neutral particles, whereas only 3\% of the electron currents (see the difference in $I_\mathrm{e,vc,side}$) change their direction. Consequently, ion-neutral particle collisions have little effect on the destination of electrons. However, note that this study ignores electron-neutral particle collisions, which may affect the electron current.

\begin{table}[hbt!]
\caption{Ion and electron current from thruster plume to different locations. }
\centering
\small
\begin{tabular}{lc|cccl|cccl}
\hline \hline
\multirow{2}{*}{Case} & $V_\mathrm{ne},$ 
    & \multicolumn{4}{c|}{Ion Current, $I_\mathrm{i}/I_\mathrm{i0}^{a,b}$} 
    & \multicolumn{4}{c}{Electron Current, $I_\mathrm{e}/I_\mathrm{e0}^{a,b}$} \\
    & V
    & $I_\mathrm{ne}$ & $I_\mathrm{th}$ & $I_\mathrm{ps}$ & $I_\mathrm{vc} (=I_\mathrm{vc,end}$+$I_\mathrm{vc,side})$ 
    & $I_\mathrm{ne}$ & $I_\mathrm{th}$ & $I_\mathrm{ps}$ & $I_\mathrm{vc} (=I_\mathrm{vc,end}$+$I_\mathrm{vc,side})$ \\ \hline
1B    & 0 & 0.00 & 0.00 & 0.01 & 0.99 & 0.39 & 0.06 & 0.26 & 0.29 \\
2B    & 0 & 0.00 & 0.00 & 0.01 & 0.99 (= 0.99 + 0.00) & 0.18 & 0.01 & 0.02 & 0.78 (= 0.59 + 0.19) \\
2B-BP & 0 & 0.00 & 0.01 & 0.02 & 0.96 (= 0.72 + 0.24) & 0.14 & 0.01 & 0.05 & 0.79 (= 0.57 + 0.22) \\
2B-NM & -5 & 0.00 & 0.02 & 0.04 & 0.95 (= 0.65 + 0.30) & 0.00 & 0.00 & 0.01 & 0.99 (= 0.80 + 0.19) \\
2B-NP & 5 & 0.00 & 0.01 & 0.01 & 0.98 (= 0.75 + 0.23) & 0.59 & 0.00 & 0.08 & 0.33 (= 0.16 + 0.17) \\
\hline \hline
\multicolumn{10}{l}{$^a$ All current values are normalized by the emitted ion or electron current ($I_\mathrm{i0}$, $I_\mathrm{e0}$), given in Table~\ref{tab:thrustercondition}.} \\
\multicolumn{10}{l}{$^b$ See Fig.~\ref{fig:elecschem} for definitions of the current to each part.} \\
\end{tabular}
\label{tab:facilityeffectcurrent}
\end{table}


\section{Sensitivity of Ion Plume Due to Thruster and Neutralizer Potential}

\begin{figure}[hbt!]
    \centering
    \includegraphics[width=\textwidth]{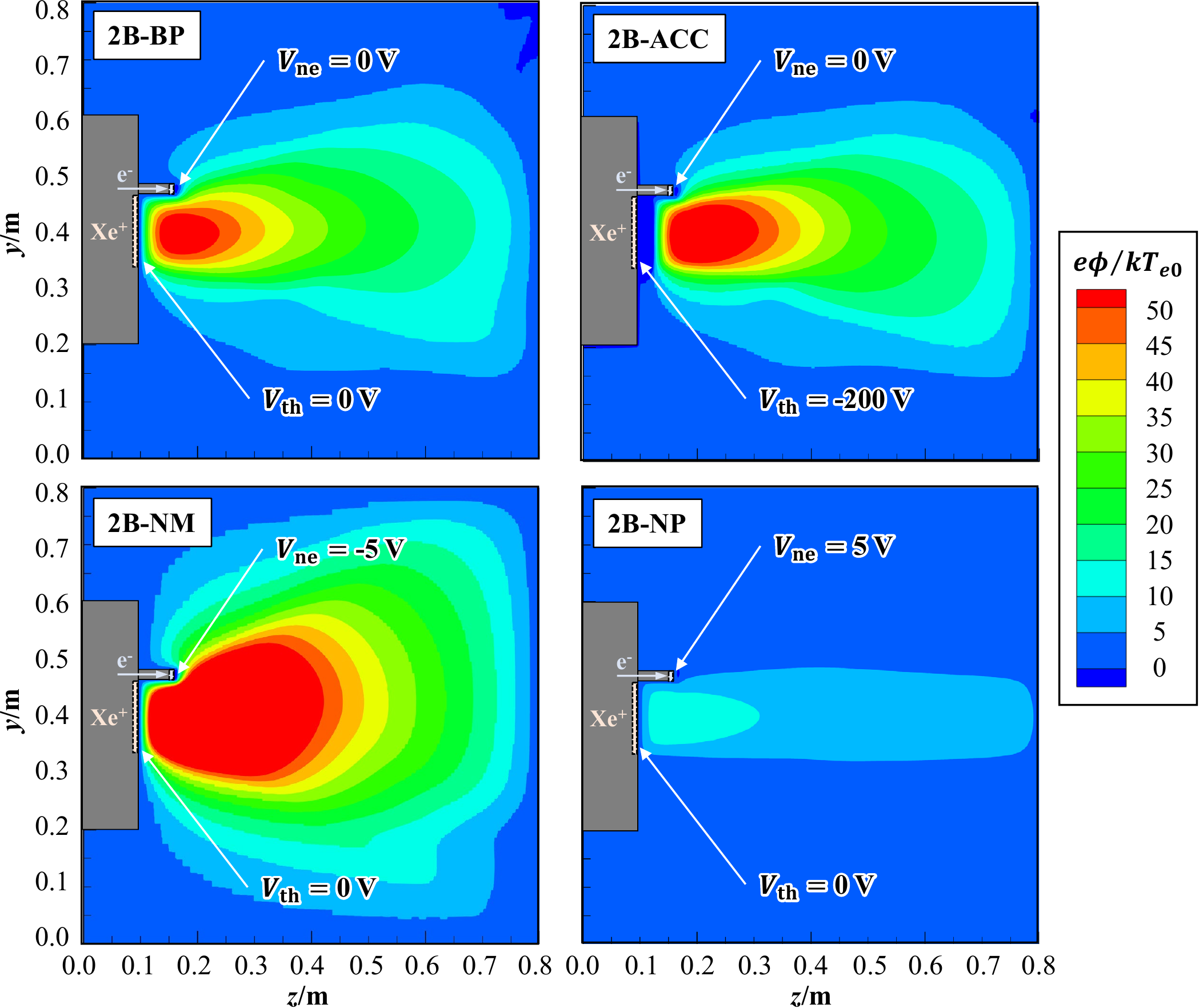}
    \caption{Electric potential in the $x=0$~m plane for various electric potential BCs.}
    \label{fig:phidist_potentialBC}
\end{figure}

This section investigates the changes in the plume due to different electric potential BCs, which can vary with each thruster potential condition. Figure~\ref{fig:phidist_potentialBC} shows the calculated potentials for the four cases shown in Table~\ref{tab:potentialcaseid}. As can be seen from this figure, the differences significantly affect the plume. A detailed discussion follows.

\subsection{Accel Grid Potential}
\label{sec:result-decelgrid}

\begin{figure}[hbt!]
    \centering
    \includegraphics[width=0.45\textwidth]{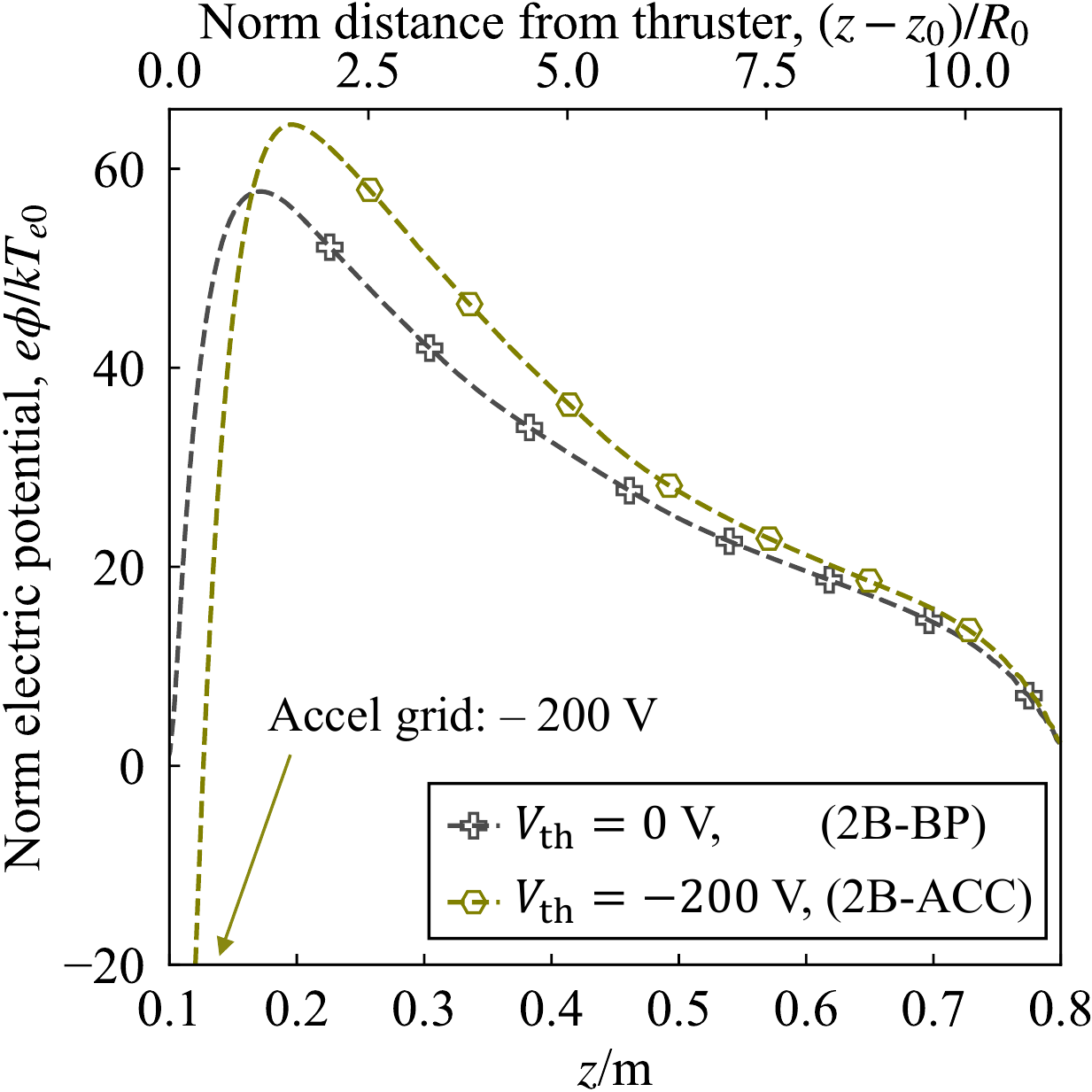}
    \caption{Electric potential along the thruster axis for different thruster exit potential BCs.}
    \label{fig:acc_phi_thruster}
\end{figure}

First, we present the results of changing the potential at the thruster exit ($V_\mathrm{th}$), simulating the case when a negatively biased accel grid is exposed to the plasma without a decel grid Case 2B-ACC. As shown in Fig.~\ref{fig:phidist_potentialBC}, low potentials are observed near the negatively biased outlet, while a higher potential region is seen downstream. In Fig.~\ref{fig:acc_phi_thruster}, the potential plotted on the thruster axis shows that in the 2B-ACC case, it is initially - 200~V but reaches a larger maximum potential than that in the 2B-BP case at $z\sim0.2$~m or 0.05~m downstream from the peak of the 2B-BP case. The reason for the larger maximum plume potential despite the lower potential of the thruster may be attributed to the degradation of the coupling between the neutralizer and the thruster.

\begin{figure}[hbt!]
    \begin{subfigure}{0.55\textwidth}
        \centering
        \includegraphics[width=\linewidth]{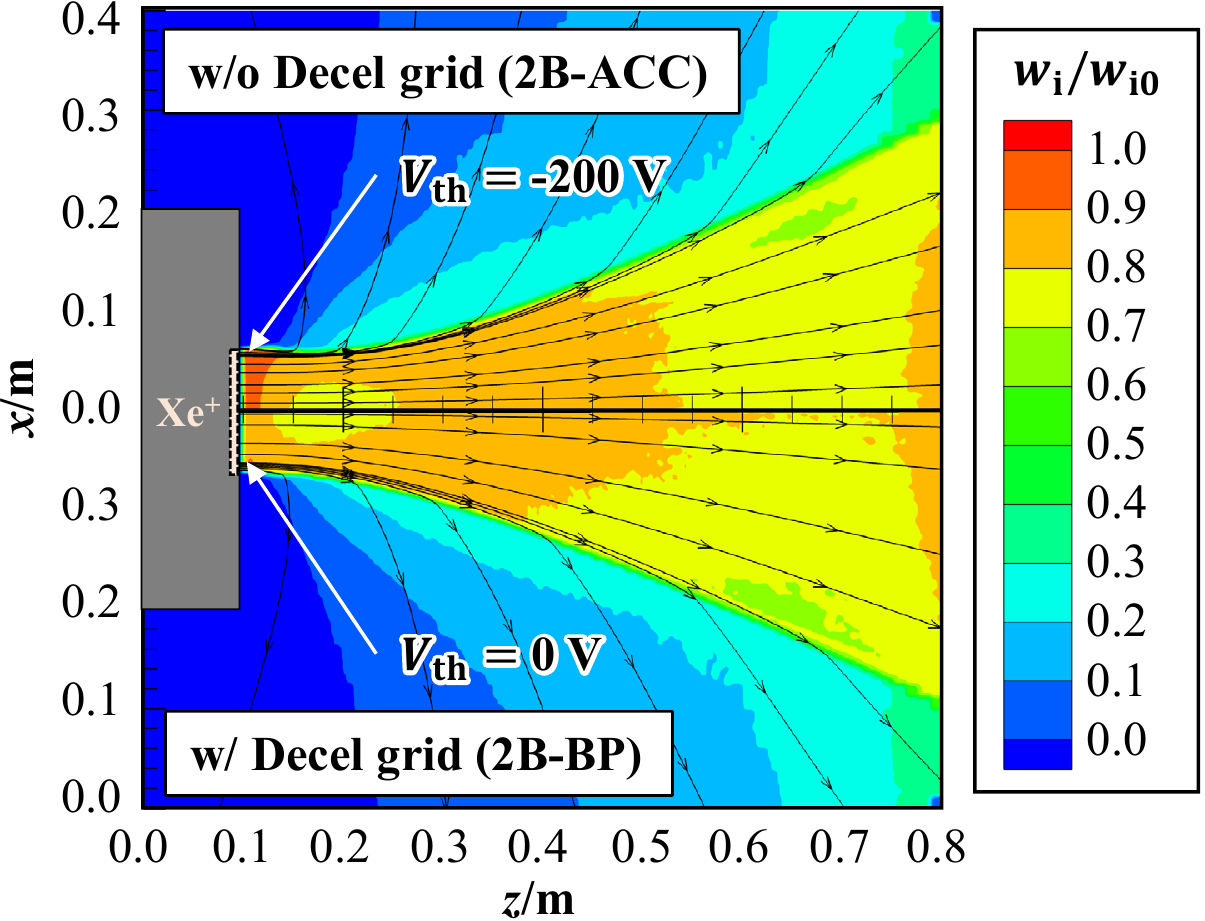}
        \caption{In the $y=0.4$~m plane} 
        \label{fig:acc_wi_zx}
    \end{subfigure}%
    \begin{subfigure}{0.45\textwidth}
        \centering
        \includegraphics[width=\linewidth]{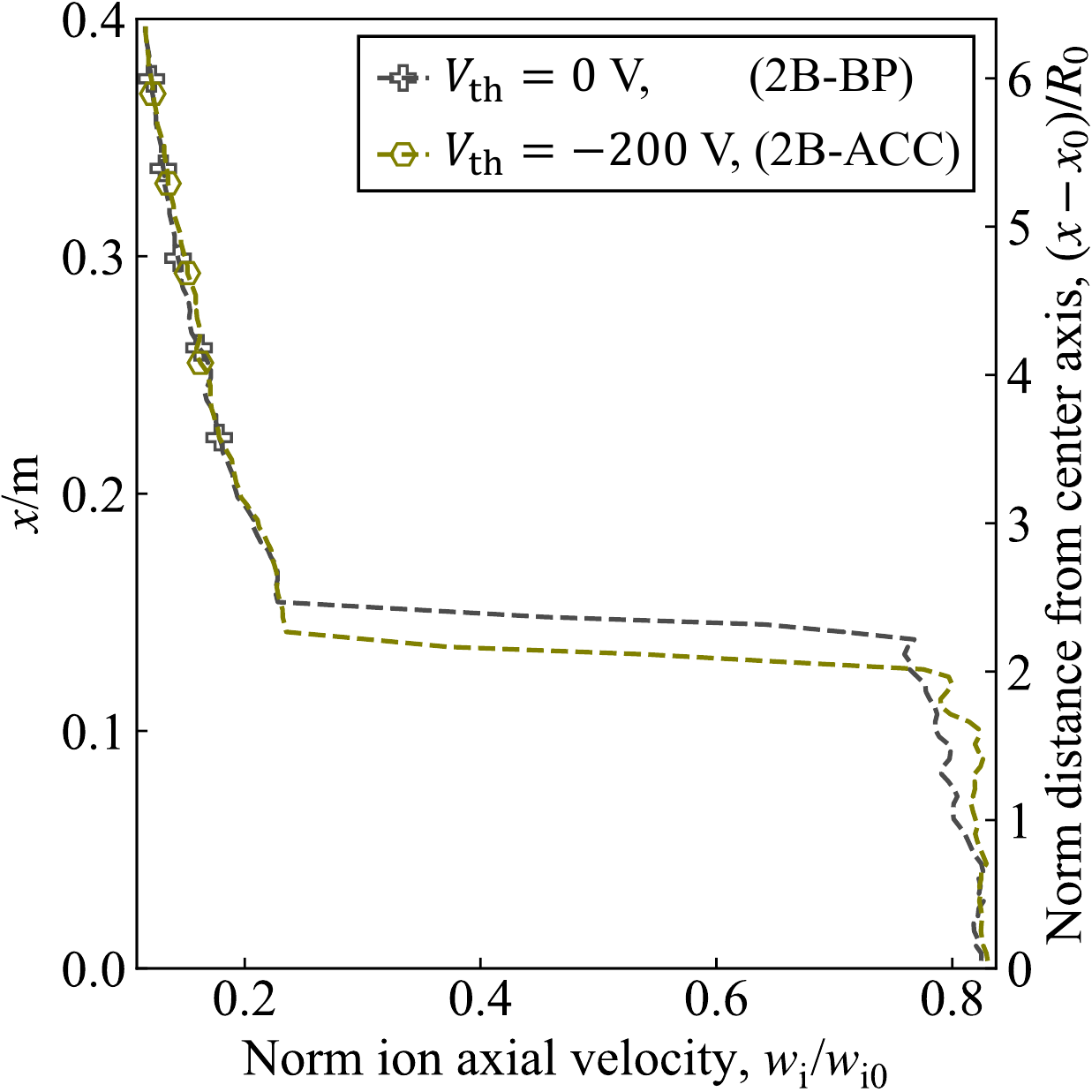}
        \caption{On the ($z$, $y$) = (0.45, 0.4) m} 
        \label{fig:acc_wi_xline}
    \end{subfigure}%
    \caption{Ion axial velocity for $V_\mathrm{th}=-200$ V vs. $V_\mathrm{th}=0$~V.}
    \label{fig:acc_wion}
\end{figure}

In the 2B-ACC case, the ion inflow velocity increases by an amount proportional to the square root of the accel grid potential of 200~V because there is no deceleration by a decel grid (2B-ACC). The axial ion velocity, $w_\mathrm{i}$, in the $y=0.4$~m plane is shown in Fig.\ref{fig:acc_wi_zx}, where $w_\mathrm{i}$ is normalized by the reference beam ion velocity of $w_\mathrm{i0}=30,0000$ m/s (612~eV). While the 2B-ACC case shows larger velocities near the thruster exit, there is almost no difference between the two cases downstream. The line plots in Fig.~\ref{fig:acc_wi_xline} also show that the velocities are almost identical, but the beam width is narrower in the 2B-ACC case. This is due to the convergence of the plume near the thruster caused by the external electric field induced by the potentials between the plasma screen of 0~V and the thruster exit of -200 V. This was also observed in an experiment (Ref.~\cite{Nguyen2021-vq}), where the grounded thruster cover around the accel grid reduced the beam divergence angle.

\subsection{Neutralizer Potential}
\label{sec:result-keepervoltage}

\begin{figure}[hbt!]
    \begin{subfigure}{0.5\textwidth}
        \centering
        \includegraphics[width=0.9\linewidth]{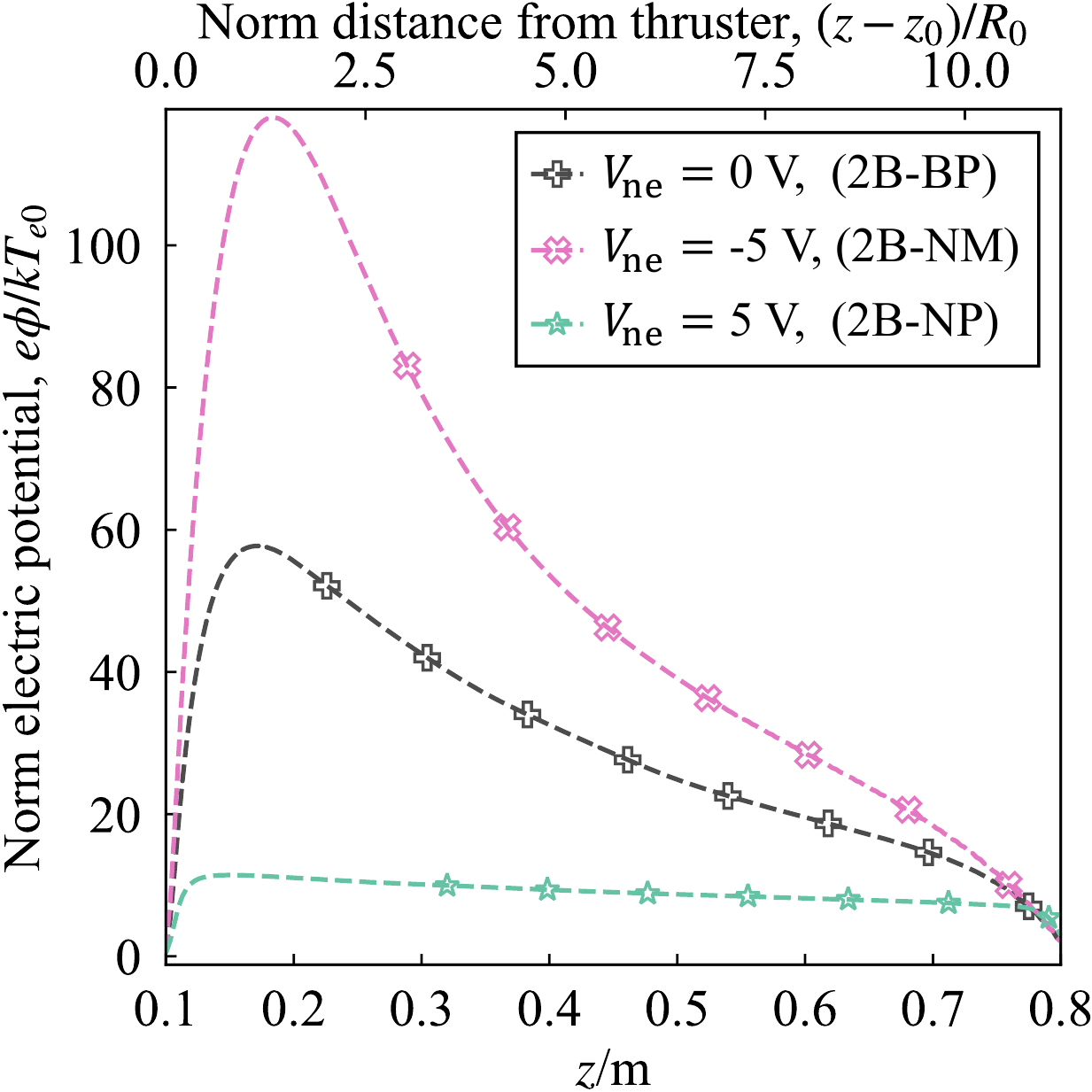}
        \caption{Along the thruster axis, entire downstream.} 
        \label{fig:neutl_phi_thruster}
    \end{subfigure}%
    \begin{subfigure}{0.5\textwidth}
        \centering
        \includegraphics[width=0.9\linewidth]{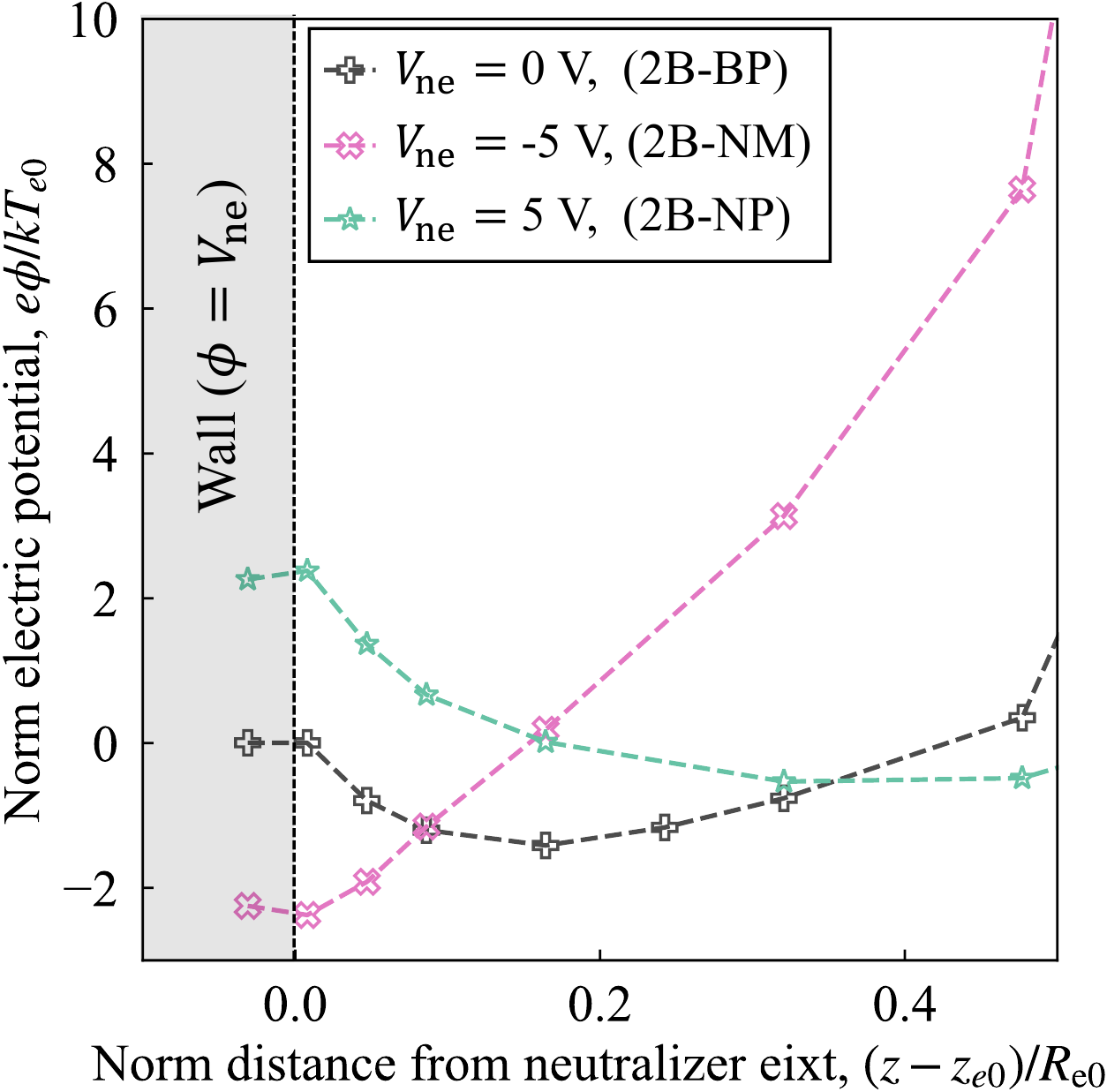}
        \caption{Along the neutralizer axis, near the exit.} 
        \label{fig:neutl_phi_cathode}
    \end{subfigure}%
    \caption{Electric potential in $z$-direction for different neutralizer exit potential BCs.}
    \label{fig:Neutl_potentialline}
\end{figure}

Next, we investigate the cases where the neutralizer exit potential ($V_\mathrm{ne}$) is 5~V lower (2B-NM) and 5~V higher (2B-NP) with respect to the ground voltage of 0~V of the 2B-BP case. As shown in Fig~\ref{fig:phidist_potentialBC}, the potential of the plasma plume changes significantly even though the exit potential is only changed by $\pm$5 V. Figure~\ref{fig:neutl_phi_thruster} shows the potential along the thruster axis where it can be seen that the maximum potential is increased by $e\phi / kT_\mathrm{e0} \sim 50$ for the 2B-NM case, while for 2B-NP it is decreased by $e\phi / kT_\mathrm{e0} \sim 50$. Additionally, as shown in Fig~\ref{fig:neutl_phi_cathode}, a comparison of the potentials near the neutralizer indicates that as the neutralizer potential is lowered, the downstream potential well becomes smaller, and it disappears completely in the 2B-NM case. The reason that this occurs is related to the current distributions, as we discuss next.

Table~\ref{tab:facilityeffectcurrent} shows the currents to different parts of the GIT for three cases (2B-BP, 2B-NM, and 2B-NP). Regarding ion currents, the large plume potential for Case 2B-NM shown in Fig~\ref{fig:phidist_potentialBC} causes more divergence of the ion beam, and a larger number of CEX ions return to the chamber side walls and the thruster, compared to case 2B-BP. However, a more significant difference is observed in the electron currents than in ion currents. As shown in Fig~\ref{fig:neutl_phi_cathode}, when $V_\mathrm{ne}$ is small, all electrons are allowed to flow downstream, while when $V_\mathrm{ne}$ is large, more electrons return to the neutralizer, as a result of virtual cathode formation. Here, $I_\mathrm{e0}-I_\mathrm{e,ne}$ can be considered an effective emission current from the neutralizer, $I_\mathrm{e,eff}$, which increases as $V_\mathrm{ne}$ decreases. This trend is consistent with experimental results showing that lowering the coupling voltage of the neutralizer releases more electron current~\cite{Yizhou2017-ee}. Furthermore, there is a large difference in $I_\mathrm{vc,end}$, but a small difference in $I_\mathrm{vc,side}$ because electrons do not return to the neutralizer.

\begin{figure}[hbt!]
    \centering
    \includegraphics[width=0.45\textwidth]{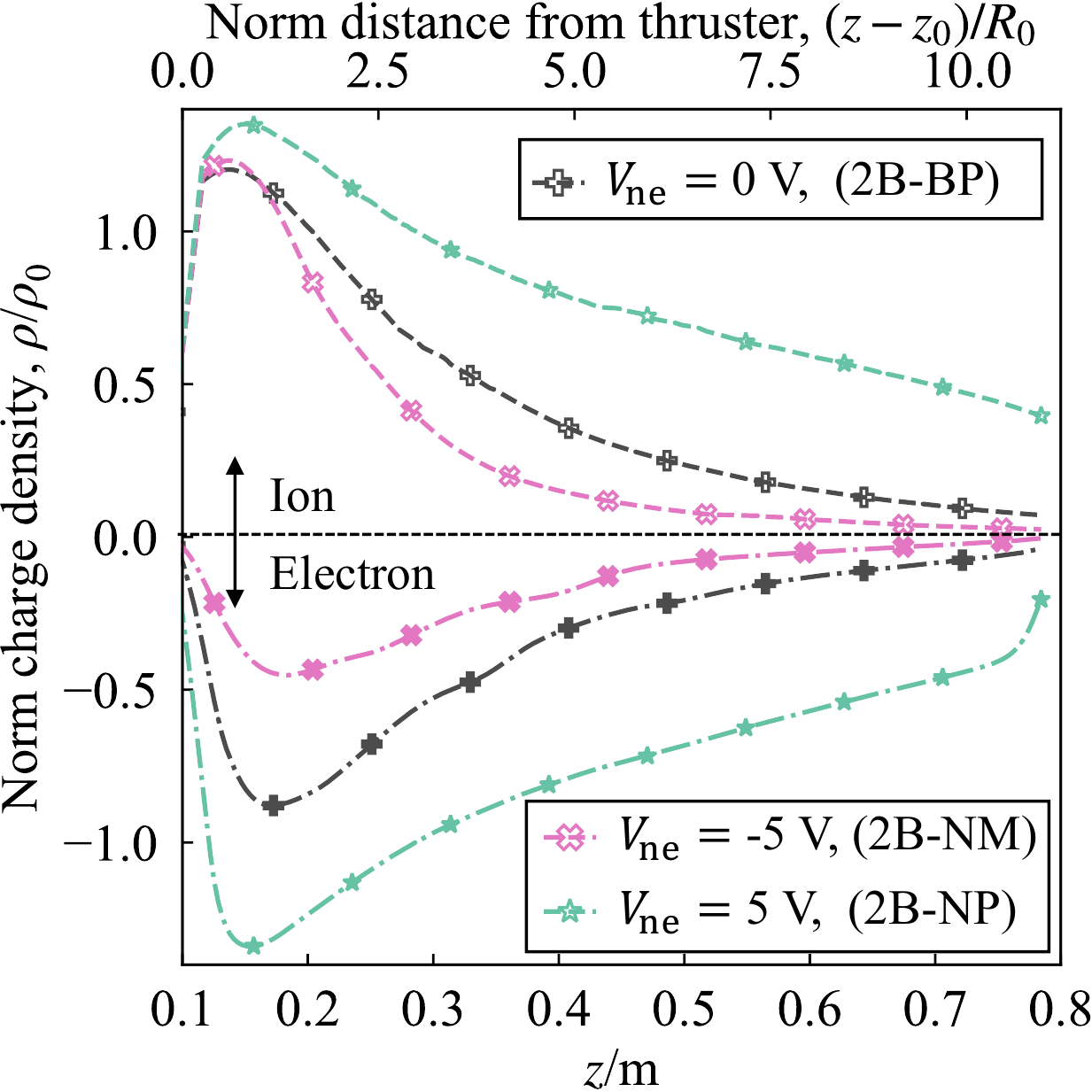}
    \caption{Volume charge density along the thruster axis for different neutralizer exit potential BCs.}
    \label{fig:neutl_rho_zline}
\end{figure}

Despite the increase in $I_\mathrm{e,eff}$, the plume has a very high potential in the 2B-NM case. The electric potential where electrons are emitted with respect to the vacuum chamber, $V_\mathrm{ne}$, changes the electron motion in the vacuum chamber. Electrons produced at lower potentials than the vacuum chamber ($V_\mathrm{ne}=-5$~V, 2B-NM) do not return to the plume because of the sheath in front of the vacuum chamber walls and are all absorbed by the wall, resulting in a shortage of electrons for neutralization. On the other hand, the sheath reflects almost all electrons produced at higher potentials ($V_\mathrm{ne}=5$~V) except for those with high kinetic energy. As a result, a sufficient number of electrons remain in the chamber when $V_\mathrm{ne}$ is large (2B-NP), as shown in Fig.~\ref{fig:neutl_rho_zline}. In Table~\ref{tab:facilityeffectcurrent}, $I_\mathrm{vc}$ decreases as $V_\mathrm{ne}$ increases, which confirms this electron confinement effect inside the vacuum chamber from the viewpoint of the currents.

\begin{figure}[hbt!]
    \centering
    \includegraphics[width=0.45\textwidth]{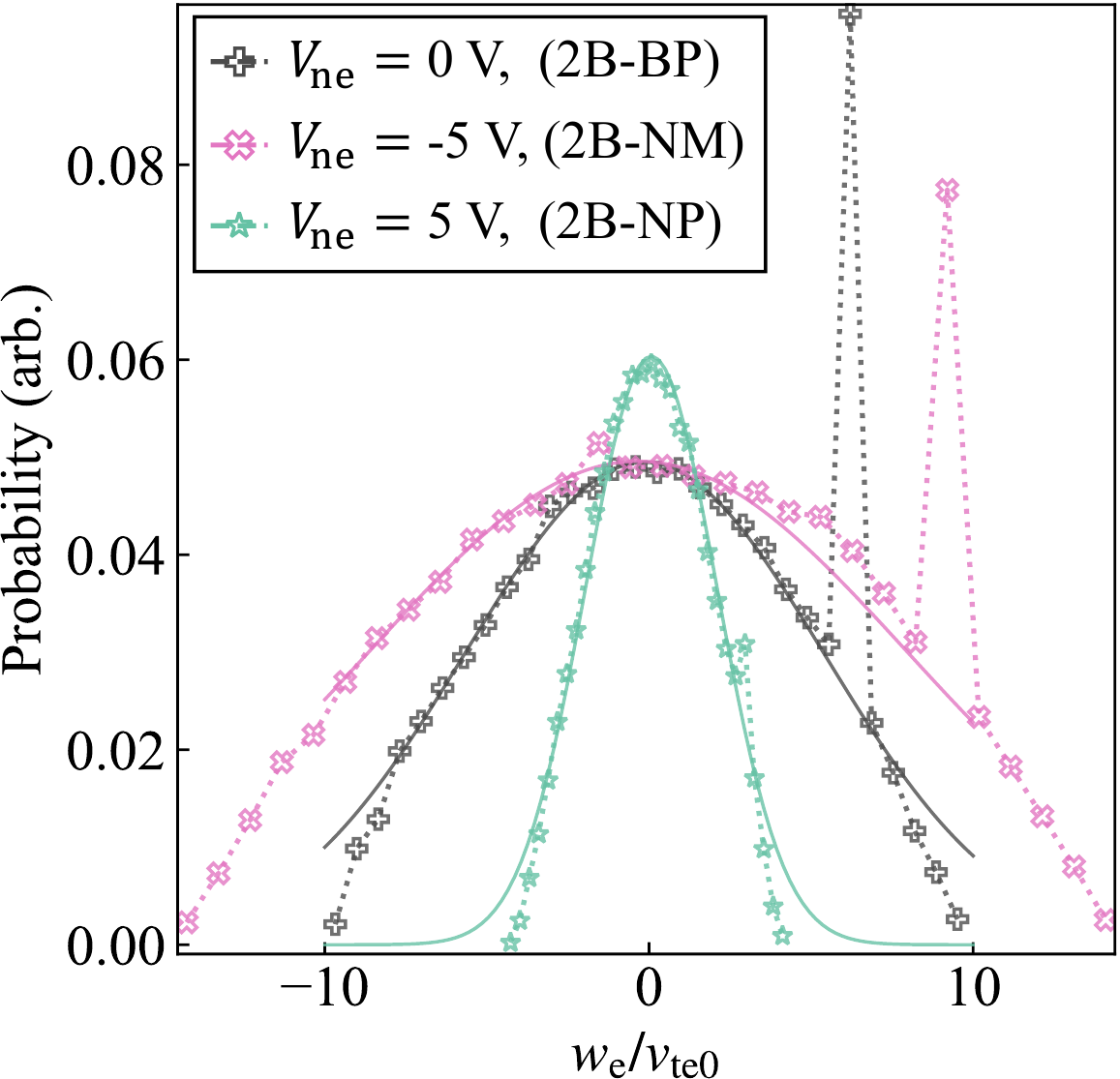}
    \caption{EVDF in $z$-direction for the 2B-BP, 2B-NM, and 2B-NP cases. Computational electrons are sampled at $R_0$ downstream from the thruster exit center. The solid lines are Maxwellian distribution fitting curves, based on Eq~(\ref{eq:maxwellian}), and the normalized electron temperature obtained by fitting, $T_\mathrm{e}/T_\mathrm{e0}$, are 15~eV (2B-BP), 34~eV (2B-NM), and 2~eV (2B-NP), where $T_\mathrm{e0}=2$ eV.}
    \label{fig:EDF_Neutl}
\end{figure}

Figure~\ref{fig:EDF_Neutl} shows the EVDFs in the $z$-directions obtained by sampling computational electrons at $R_0$ downstream on the thruster axis for the 2B-BP, 2B-NM, and 2B-NP cases. Two distributions are clearly seen in the cases 2B-BP and 2B-NM, but almost all electrons are thermalized in the 2B-NP case. Fitting the electron temperature to the distribution for the population centered around zero velocity reveals that $T_\mathrm{e}$ decreases as $V_\mathrm{ne}$ increases. This is because the plume potential is reduced by the above-mentioned change in electron motion (see Figs.~\ref{fig:phidist_potentialBC} and \ref{fig:neutl_phi_thruster}), allowing the otherwise-trapped electrons to move with less energy.

\section{Conclusions}

In this article, we have simulated ion thruster plumes in a vacuum chamber and space configurations using the in-house multi-GPU CHAOS solver. A fully kinetic PIC-DSMC approach is applied to model electron motion emitted from two types of external neutralizers separated from the thruster. This study qualitatively confirms important aspects of electrical facility effects and improves our understanding of how electrons move in a vacuum test chamber for practical GIT configurations.

First, we have shown that the plume potential and electron temperature are larger in the ground test than in the space environment because the former absorbs all the electrons, reducing the number of electrons trapped in the ion beam potential. This causes a large change in the electron current. The backflow current from the plume to the propulsion system, including the neutralizer, obtained under vacuum chamber conditions is less than half that under space conditions, indicating that most of the current flows to the downstream wall. Another facility effect that we investigated is how background pressure affects the plume. In particular, it was shown that the electron flow is also changed when slow CEX ions accumulate in areas with negative charge density. This causes the maximum potential of the plume and electron temperature to decrease. As for the ion current, the reverse current to the thruster and the current to the side walls of the chamber increase significantly. The electron current also changes but at a smaller rate than the ion current.

For further practical insight, we investigated the coupling between the ion beam and electrons by changing the position of the neutralizer while keeping the emitted electron current constant. As the neutralizer is moved downstream from the thruster wall, the maximum potential of the plume decreases, and the minimum potential of the virtual cathode created at the neutralizer exit increases. Electrons emitted from the neutralizer flow downstream in the central plane of the thruster, moving back and forth across the center of the plume where the potential is high, creating an asymmetric charge density distribution.

In addition, we have also simulated different electric potentials at the ion and electron source exits. In the case of no decel grid, where the -200~V accel grid is exposed to the plasma, the maximum potential increased due to the difference in coupling with the electron source, regardless of the lower exit potential. From a performance standpoint, the ion beam is more focused, but the final ion velocity remains the same. When the neutralizer potential was lower than the chamber wall, the plume potential increased significantly because more electrons were extracted and absorbed toward the high potential chamber wall, resulting in insufficient neutralization. Conversely, when the potential of the neutralizer is higher than the chamber, the chamber acts as a cage for electrons, and the plume is more neutralized.


\section*{Acknowledgments}
This work was partially supported by NASA through the Joint Advanced Propulsion Institute, a NASA Space Technology Research Institute, grant number 80NSSC21K1118. This work used Delta at the National Center for Supercomputing Applications through allocation TG-PHY220010 from the Advanced Cyberinfrastructure Coordination Ecosystem: Services \& Support (ACCESS) program, which is supported by National Science Foundation grants \#2138259, \#2138286, \#2138307, \#2137603, and \#2138296.

\section*{References}

\bibliography{reference}

\begin{thebibliography}{10}

\bibitem{Holste2020-gr}
K~Holste, P~Dietz, S~Scharmann, K~Keil, T~Henning, D~Zsch{\"a}tzsch,
  M~Reitemeyer, B~Nausch{\"u}tt, F~Kiefer, F~Kunze, J~Zorn, C~Heiliger,
  N~Joshi, U~Probst, R~Th{\"u}ringer, C~Volkmar, D~Packan, S~Peterschmitt, K-T
  Brinkmann, H-G Zaunick, M~H Thoma, M~Kretschmer, H~J Leiter, S~Schippers,
  K~Hannemann, and P~J Klar.
\newblock Ion thrusters for electric propulsion: Scientific issues developing a
  niche technology into a game changer.
\newblock {\em The Review of scientific instruments}, 91(6):061101, June 2020.

\bibitem{Kerslake1993-my}
William~R Kerslake and Louis~R Ignaczak.
\newblock Development and flight history of the {SERT} {II} spacecraft.
\newblock {\em Journal of Spacecraft and Rockets}, 30(3):258--290, May 1993.

\bibitem{Kerslake1982-mj}
W~R Kerslake and L~R Ignaczak.
\newblock {SERTII1979-1981} tests: Plasma thrust and neutralizer measurements.
\newblock {\em Journal of Spacecraft and Rockets}, 19(3):236--240, May 1982.

\bibitem{Nakayama2015-ew}
Yoshinori Nakayama and Futoshi Tanaka.
\newblock Experimental visualization of ion thruster neutralization phenomena.
\newblock {\em IEEE transactions on plasma science}, 43(1):269--276, January
  2015.

\bibitem{Dale2020-dd}
Ethan Dale, Benjamin Jorns, and Alec Gallimore.
\newblock Future directions for electric propulsion research.
\newblock {\em Aerospace}, 7(9):120, August 2020.

\bibitem{Li2019-eo}
Wen-Bo Li, Hong Li, Yong-Jie Ding, Li-Qiu Wei, Qian Gao, Shi-Lin Yan, Tian-Hang
  Meng, Xi-Ming Zhu, and Da-Ren Yu.
\newblock Study on electrons conduction paths in hall thruster ignition
  processes with the cathode located inside and outside the magnetic
  separatrix.
\newblock {\em Acta astronautica}, 155:153--159, February 2019.

\bibitem{Polansky2013-nv}
John Polansky, Joseph Wang, and Ning Ding.
\newblock Experimental investigation on plasma plume potential.
\newblock {\em IEEE transactions on plasma science}, 41(12):3438--3447,
  December 2013.

\bibitem{Conde2022-zp}
L~Conde, P~E Maldonado, J~Damba, J~Gonzalez, J~L Domenech-Garret, J~M Donoso,
  and M~A Castillo.
\newblock Physics of the high specific impulse alternative low power hybrid ion
  engine (alphie): Direct thrust measurements and plasma plume kinetics.
\newblock {\em Journal of applied physics}, 131(2):023302, January 2022.

\bibitem{Zhao2018-kq}
Yinjian Zhao, Joseph Wang, and Hideyuki Usui.
\newblock Simulations of ion thruster beam neutralization using a
  {Particle--Particle} model.
\newblock {\em Journal of Propulsion and Power}, 34(5):1109--1115, September
  2018.

\bibitem{Hu2015-mr}
Yuan Hu and Joseph Wang.
\newblock Electron properties in collisionless mesothermal plasma expansion:
  Fully kinetic simulations.
\newblock {\em IEEE transactions on plasma science}, 43(9):2832--2838,
  September 2015.

\bibitem{Wang2012-hy}
Joseph Wang, Ouliang Chang, and Yong Cao.
\newblock {Electron--Ion} coupling in mesothermal plasma beam emission: Full
  particle {PIC} simulations.
\newblock {\em IEEE transactions on plasma science}, 40(2):230--236, February
  2012.

\bibitem{Hu2020-vs}
Yuan Hu, Joseph Wang, and Quanhua Sun.
\newblock Geometrically self-similar ion acceleration in collisionless plasma
  beam expansion.
\newblock {\em Plasma Sources Science and Technology}, 29(12):125004, December
  2020.

\bibitem{Wang2015-pm}
Joseph Wang, Daoru Han, and Yuan Hu.
\newblock Kinetic simulations of plasma plume potential in a vacuum chamber.
\newblock {\em IEEE transactions on plasma science}, 43(9):3047--3053,
  September 2015.

\bibitem{Li2019-xn}
Min Li, Mario Merino, Eduardo Ahedo, and Haibin Tang.
\newblock On electron boundary conditions in {PIC} plasma thruster plume
  simulations.
\newblock {\em Plasma Sources Science and Technology}, 28(3):034004, March
  2019.

\bibitem{Wang2019-gy}
Joseph Wang and Yuan Hu.
\newblock On the limitations of hybrid particle-in-cell for ion thruster plume
  simulations.
\newblock {\em Physics of plasmas}, 26(10):103502, October 2019.

\bibitem{Nuwal2020-pa}
Nakul Nuwal, Revathi Jambunathan, and Deborah~A Levin.
\newblock Kinetic modeling of spacecraft surfaces in a plume backflow region.
\newblock {\em IEEE transactions on plasma science}, 48(12):4305--4325,
  December 2020.

\bibitem{Jambunathan2020-io}
Revathi Jambunathan and Deborah~A Levin.
\newblock Kinetic, 3-d, {PIC-DSMC} simulations of ion thruster plumes and the
  backflow region.
\newblock {\em IEEE transactions on plasma science}, 48(6):2017--2034, June
  2020.

\bibitem{Nishii2023-ju}
Keita Nishii and Deborah~A Levin.
\newblock {Three-Dimensional} kinetic simulation of an ion thruster plume with
  carbon backsputtering in a vacuum chamber.
\newblock preprint, ArXivID: 2303.07496, Mar. 2023 (available at
  https://doi.org/10.48550/arXiv.2303.07496).

\bibitem{Usui2013-al}
Hideyuki Usui, Akihiko Hashimoto, and Yohei Miyake.
\newblock Electron behavior in ion beam neutralization in electric propulsion:
  full particle-in-cell simulation.
\newblock {\em Journal of physics. Conference series}, 454(1):012017, August
  2013.

\bibitem{Jambunathan2018-xf}
Revathi Jambunathan and Deborah~A Levin.
\newblock {CHAOS}: An octree-based {PIC-DSMC} code for modeling of electron
  kinetic properties in a plasma plume using {MPI-CUDA} parallelization.
\newblock {\em Journal of computational physics}, 373:571--604, November 2018.

\bibitem{Brieda2018-wj}
Lubos Brieda.
\newblock Model for {Steady-State} fully kinetic ion beam neutralization
  studies.
\newblock {\em IEEE transactions on plasma science}, 46(3):556--562, March
  2018.

\bibitem{Brieda2005-kj}
Lubos Brieda and Joseph Wang.
\newblock Modelling ion thruster beam neutralization.
\newblock In {\em 41st {AIAA/ASME/SAE/ASEE} Joint Propulsion Conference \&
  Exhibit}, Joint Propulsion Conferences, Reston, Virigina, July 2005. American
  Institute of Aeronautics and Astronautics.

\bibitem{Jambunathan2020-yx}
Revathi Jambunathan and Deborah~A Levin.
\newblock A {Self-Consistent} open boundary condition for fully kinetic plasma
  thruster plume simulations.
\newblock {\em IEEE transactions on plasma science}, 48(3):610--630, March
  2020.

\bibitem{Birdsall1991-lh}
C~K Birdsall.
\newblock Particle-in-cell charged-particle simulations, plus monte carlo
  collisions with neutral atoms, {PIC-MCC}.
\newblock {\em IEEE transactions on plasma science}, 19(2):65--85, April 1991.

\bibitem{Bird1987-ba}
G~A Bird.
\newblock Direct simulation of high-vorticity gas flows.
\newblock {\em Physics of fluids}, 30(2):364, 1987.

\bibitem{Foster2022-ck}
John~E Foster, Tyler Topham, and Andria Sperry.
\newblock Review of facility effects on gridded ion thruster operation and
  performance.
\newblock In {\em Proceedings of the 37th International Electric Propulsion
  Conference}, IEPC-2022-279, June 2022.

\bibitem{Wirz2008-fj}
Richard~E Wirz, John~R Anderson, Dan~M Goebel, and Ira Katz.
\newblock Decel grid effects on ion thruster grid erosion.
\newblock {\em IEEE transactions on plasma science}, 36(5):2122--2129, October
  2008.

\bibitem{Araki2013-em}
Samuel~J Araki and Richard~E Wirz.
\newblock {Ion--Neutral} collision modeling using classical scattering with
  {Spin-Orbit} free interaction potential.
\newblock {\em IEEE transactions on plasma science}, 41(3):470--480, March
  2013.

\bibitem{Miller2002-ne}
J~Scott Miller, Steve~H Pullins, Dale~J Levandier, Yu-Hui Chiu, and Rainer~A
  Dressler.
\newblock Xenon charge exchange cross sections for electrostatic thruster
  models.
\newblock {\em Journal of applied physics}, 91(3):984--991, February 2002.

\bibitem{Serikov1999-jq}
V~V Serikov, S~Kawamoto, and K~Nanbu.
\newblock Particle-in-cell plus direct simulation monte carlo ({PIC-DSMC})
  approach for self-consistent plasma-gas simulations.
\newblock {\em IEEE transactions on plasma science}, 27(5):1389--1398, October
  1999.

\bibitem{Korkut2017-zq}
Burak Korkut, Deborah~A Levin, and Ozgur Tumuklu.
\newblock Simulations of ion thruster plumes in ground facilities using
  adaptive mesh refinement.
\newblock {\em Journal of Propulsion and Power}, 33(3):681--696, May 2017.

\bibitem{Kitamura2007-sk}
Shoji Kitamura, Yasushi Ohkawa, Yukio Hayakawa, Hideki Yoshida, and Katsuhiro
  Miyazaki.
\newblock Overview and research status of the {JAXA} 150-mn ion engine.
\newblock {\em Acta astronautica}, 61(1):360--366, June 2007.

\bibitem{Perales-Diaz2021-ci}
Jes{\'u}s Perales-D{\'\i}az, Filippo Cichocki, Mario Merino, and Eduardo Ahedo.
\newblock Formation and neutralization of electric charge and current of an ion
  thruster plume.
\newblock {\em Plasma Sources Science and Technology}, 30(10):105023, October
  2021.

\bibitem{Miyasaka2012-cy}
Takeshi Miyasaka, Katsuo Asato, Fakhuradzi~Bin Baharudin, Hitoshi Sugiyama, and
  Ikkoh Funaki.
\newblock Study on electron distributions in a three dimensional particle
  simulation of an ion engine.
\newblock {\em Transactions of the Japan Society for Aeronautical and Space
  Sciences, Aerospace Technology Japan}, 10(ists28):Pb\_13--Pb\_17, 2012.

\bibitem{Wirz2003-gd}
Richard Wirz, Daniel Goebel, Colleen Marrese, and Juergen Mueller.
\newblock Development of cathode technologies for a miniature ion thruster.
\newblock In {\em 39th {AIAA/ASME/SAE/ASEE} Joint Propulsion Conference and
  Exhibit}, Joint Propulsion Conferences, Reston, Virigina, July 2003. American
  Institute of Aeronautics and Astronautics.

\bibitem{Kottke2019-qa}
Nils~Gerrit Kottke, Max Vaupel, Martin Tajmar, Werner Konrad, Noah Saks, and
  Franz~Georg Hey.
\newblock Comparison of the thermionic emission properties of {LaB6} and
  {C12A7}.
\newblock In {\em Proceedings of the 36th International Electric Propulsion
  Conference, Vienna, Austria}, pages 15--20, 2019.

\bibitem{Yizhou2017-ee}
Jin Yizhou, Juan Yang, Sun Jun, Liu Xianchuang, and Yizhi Huang.
\newblock Experiment and analysis of the neutralization of the electron
  cyclotron resonance ion thruster.
\newblock {\em Plasma Science and Technology}, 19(10):105502, August 2017.

\bibitem{Bellissimo2020-qf}
Alessandra Bellissimo, Gian~Marco Pierantozzi, Alessandro Ruocco, Giovanni
  Stefani, Olga~Yu Ridzel, Vytautas Asta{\v s}auskas, Wolfgang S~M Werner, and
  Mauro Taborelli.
\newblock Secondary electron generation mechanisms in carbon allotropes at low
  impact electron energies.
\newblock {\em Journal of Electron Spectroscopy and Related Phenomena},
  241:146883, May 2020.

\bibitem{Ward1968-dc}
J~W Ward and H~J King.
\newblock Mercury hollow cathode plasma bridge neutralizers.
\newblock {\em Journal of Spacecraft and Rockets}, 5(10):1161--1164, October
  1968.

\bibitem{Hershkowitz2005-la}
Noah Hershkowitz.
\newblock Sheaths: More complicated than you think.
\newblock {\em Physics of plasmas}, 12(5):055502, May 2005.

\bibitem{Nguyen2021-vq}
Huong T~T Nguyen, Hoai-Duc Vu, and Jichul Shin.
\newblock Evaluation of ion beam behavior in 50 {W} class {RF} ion thruster.
\newblock {\em International Journal of Aerospace Engineering}, 2021, September
  2021.

\end{thebibliography}
\bibliographystyle{unsrt}

\end{document}